\documentclass[a4paper,11pt]{article}
\usepackage{jheppub} 
\usepackage{lineno}

\usepackage{graphics}%
\usepackage[mathcal]{euscript}
\usepackage[latin1]{inputenc}
\usepackage{latexsym}
\usepackage{hyperref}
\usepackage{amsmath}    
\usepackage{graphicx}   
\usepackage{verbatim}      
\usepackage{subfigure}  
\usepackage{hyperref}   
\usepackage{bm}
\usepackage{graphicx}
\usepackage{amssymb}
\usepackage{bbm}
\usepackage{float}
\usepackage{slashed}
\usepackage{amsfonts}
\usepackage{euscript}
\usepackage{amsmath}    
\usepackage{mathrsfs} 
\usepackage{bbding}
\usepackage{pifont}
\usepackage{cancel}
\usepackage{multirow}
\usepackage{soul}
\usepackage{physics}
\usepackage{eufrak}
\usepackage{breqn}
\usepackage{tikz}

\usepackage{xcolor}

\newcommand{\q}[1]{\mathcal Q\pqty{#1}}
\newcommand{\dif}[1]{\mathbb D\pqty{#1}}

\makeindex

\begin{document}


\title{Inner horizon instability via the trace anomaly effective action}

\author{Julio Arrechea,}
\emailAdd{julio.arrechea@sissa.it}
\author{Giulio Neri and}
\emailAdd{gneri@sissa.it}
\author{Stefano Liberati}
\emailAdd{liberati@sissa.it}
\affiliation{SISSA, Via Bonomea 265, 34136 Trieste, Italy and INFN Sezione di Trieste}
\affiliation{IFPU - Institute for Fundamental Physics of the Universe, Via Beirut 2, 34014 Trieste, Italy}

\abstract{In quantum field theory applied to black hole spacetimes, substantial evidence suggests that the Unruh and Hartle-Hawking vacuum states become singular at Cauchy horizons. This raises essential questions regarding the impact of quantum field backreaction on the stability of Cauchy horizons in static scenarios and inner horizons in evolving spacetimes. To approach this problem, we employ analytic approximations to the renormalized stress-energy tensor (RSET) of quantum fields in four dimensions. Specifically, we utilize the anomaly-induced effective action, which generates four-dimensional approximate RSETs through a pair of auxiliary scalar fields that satisfy higher-order equations of motion. The boundary conditions imposed on these auxiliary fields yield RSETs with leading-order terms that mimic the behaviour of different vacuum states.
This study presents the first application of the anomaly-induced effective action method to Reissner-Nordstr\"om black hole interiors, evaluating its accuracy, applicability, and connections with prior RSET approximations. Among the range of possible states accessible through this method, we found none that remain regular at both the event and Cauchy horizons, aligning with theoretical expectations. The method shows strong agreement with exact four-dimensional RSET results for the Hartle-Hawking state but does not fully capture the unique characteristics of the Unruh state in Reissner-Nordstr\"om spacetimes. We conclude by suggesting possible extensions to address these limitations.}



\maketitle

\tableofcontents
\section{Introduction}

The theory of general relativity (GR) predicts that the formation of trapped regions and spacetime singularities is inevitable in standard gravitational collapse~\cite{Penrose1965, Hawking1970}. In the simplest situation described by the collapse of a non-rotating, chargeless dust sphere~\cite{OppenheimerSnyder1939}, the spacetime geometry quickly approaches the Schwarzschild vacuum solution as the influence of the matter initiating the collapse decreases with time~\cite{Price:1971fb,Price:1972pw}. The resulting curvature singularity poses no predictability problems for the theory since it is shielded by an event horizon, thus bearing no causal influence on outside observers. 

However, this is no longer the case when quantum physics enters the picture. Quantum fields that start in the vacuum state in the asymptotic past (the \textit{in} state~\cite{FabbriNavarro-Salas2005}) are affected by the dynamics of the collapsing star and evolve towards the Unruh state at late times~\cite{Davies:1976ei,ParentaniPiran1994,Arderucio-Costa:2017etb,Juarez-Aubry:2018ofz,Balbinot:2023vcm}. Some time after the formation of the horizon, according to semiclassical calculations, distant observers would perceive an outflux of thermal radiation at the Hawking temperature, accompanied by a negative energy flux across the horizon which makes the trapped region shrink with time~\cite{Hawking1976}. Albeit the complete disappearance of the trapped region would encompass very long timescales (all black holes are eternal from an astrophysical perspective), the theoretical tensions raised by Hawking evaporation still lack truly a satisfactory resolution to this date~\cite{Mathur2009,Curiel2010}.

It has been recently realized that the above semiclassical picture might break down in the presence of charge or rotation --- and, just as importantly, in regular black holes models~\cite{Carballo-Rubioetal2019} --- since these geometries, when stationary, all exhibit Cauchy horizons, where the energy content associated with the Unruh state has been shown to diverge~\cite{Hiscock1977, BirrellDavies1978, Hiscock:1980wr,Hollands:2019whz, LeviOri2016, Zilberman:2019buh,Hollands:2020qpe,Zilbermanetal2022,McMaken:2023uue, Zilberman:2024jns}. 
Contrarily to outer trapping horizons, which are rather oblivious to their environments, inner horizons are extremely sensitive to the local matter content, making them susceptible to a series of instabilities. Already at the classical level, Cauchy horizons suffer from classical mass inflation which is expected to lead to a singularity~\cite{PoissonIsrael1989,Ori1991,HodPiran1998,Dafermos:2012np,Dafermos:2017dbw}.

The same sort of instability was noticed in stationary regular black holes around inner horizons~\cite{Brown:2011tv,Frolov:2017rjz,Carballo-Rubio:2018pmi,Carballo-Rubio:2021bpr,Bonanno:2020fgp} (with the exception of geometries with zero surface gravity at the inner horizon~\cite{Carballo-Rubio:2022kad,Franzin:2022wai}). In these cases the avoidance of a singularity is postulated, but it is clear that the effect would rapidly bring the regular black hole into a regime where backreaction cannot be neglected and gravitational dynamics would have to be taken into account (see also~\cite{Barcelo:2022gii,DiFilippo:2022qkl,Bonanno:2022jjp,Carballo-Rubio:2023kam,Bonanno:2023qhp}).
While classical mass inflation in a stationary geometry could be seen as a manifestation of the strong cosmic censorship, it has been shown that an exponential energy accumulation would also destabilize a slowly evolving inner horizon~\cite{Carballo-Rubio:2024dca}, rendering the issue of their stability even more pressing.

At the quantum level, the aforementioned divergence of the Unruh state at Cauchy horizons would make the background spacetime inconsistent with its own vacuum state. Recent analyses suggest that the semiclassical backreaction would also tend to transform the Cauchy horizon into a curvature singularity~\cite{Zilbermanetal2022,McMaken:2024fvq,Klein:2024sdd}.
Of course, as for mass inflation, one could interpret these results as a sign that the backreaction would start being relevant for the inner horizon evolution before the formation of the Cauchy horizon. Since the divergence would never be strictly reached, the evolution of the background would have to be controlled by a combination of (perhaps competing) classical and semiclassical instabilities~\cite{Boyanov:2022xfw}.

Although the complete evolution of realistic trapped regions remains an open question, partial studies reveal an astoundingly rich family of physical phenomena~\cite{Barcelo:2020mjw,Barcelo:2022gii,Barenboim:2024Dko}. Among the possible outcomes of such processes, which we do not address in this paper, semiclassical effects suggest the possibility that trapped regions may even disappear from the inside out through an explosion of the inner horizon on very short timescales~\cite{Arrechea:2023hmo}, leaving behind either an extremal remnant or a horizonless object. Understanding the evolution of quantum fields inside the trapped region and thus how the vacuum effects influence the evolution of inner horizons is thus crucial from a phenomenological perspective. 

A first, preliminary step towards this end-goal requires determining the energy content associated with the Hartle-Hawking and Unruh vacuum states in eternal spacetimes with Cauchy horizons. 
The quantity that informs about the backreaction from quantum fields is the expectation value of the corresponding renormalized stress-energy tensor (RSET). 
Unfortunately, while this can be easily calculated on the most relevant $1+1$ dimensional black hole geometries, we do still lack simple global analytical solutions in four dimensional settings.

In eternal black hole spacetimes, the most commonly analyzed states are the Boulware state~\cite{Boulware1974}, which is singular at the future event horizon, and the Hartle-Hawking and Unruh states~\cite{Unruh1976}, which are regular instead. 
A state is called \textit{regular} (\textit{singular}) according to whether a freely-falling observer measures finite (diverging) components in the RSET.\footnote{The irregularity of a state is linked to the behaviour of the Fourier mode basis in which the field is decomposed~\cite{FabbriNavarro-Salas2005}. Perhaps the sole exception to this is the four-dimensional extremal black hole case, where the Boulware state is believed to be regular~\cite{Anderson:1995fw,Arrechea:2024cnv}.}

Exact results for the RSET of scalar fields in 3+1 dimensions in the Hartle-Hawking and Unruh states were obtained via state subtraction methods at the Cauchy horizon in both Reissner-Nordstr\"om~\cite{Zilbermanetal2022} and Kerr spacetimes~\cite{Zilberman:2022aum,Zilberman:2024jns}. The authors found that the RSET has constant flux components at the Cauchy horizon, indicating the presence of a divergence. While state subtraction works well to estimate these local divergences, calculating the RSET in the entire spacetime requires regularisation techniques such as point-splitting~\cite{Christiensen1976}, pragmatic mode-sum regularization~\cite{LeviOri2016}, or extended coordinates regularisation~\cite{TaylorBreen2022} (notice that these methods do not have an obvious extension to non-stationary spacetimes), together with the accurate numerical computation of a large number of modes. 

Once the RSET has been obtained on these backgrounds, it can be used as an $\order{\hbar}$ correction to the classical spacetime via the semiclassical Einstein equations
\begin{equation}\label{Eq:SemiEinstein}
G_{\mu\nu}=8\pi G_N\left(T^{\text{cl}}_{\mu\nu}+\langle\hat{T}_{\mu\nu}\rangle\right),
\end{equation}
where $T^{\text{cl}}_{\mu\nu}$ is the stress-tensor generating the background classical solution and $\langle\hat{T}_{\mu\nu}\rangle$ is the expectation value of the RSET on that background. Nonetheless, it is clear from the discussion above that computing the latter involves several technically challenging steps which so far have hampered a full investigation of the semiclassical backreaction in four dimensions.

Furthermore, in regimes where the RSET overcomes its $\order{\hbar}$ suppression, as in the proximity of the inner horizon of evaporating black holes, the background solution should not be trusted any more, as it will be drastically modified by quantum effects. In said situations, the validity of the semiclassical approximation as a whole is at  stake~\cite{Simon1990,FlanaganWald1996,HuVerdaguer2020,Klein:2023urp}. 
Despite this, it might be possible to gain valuable insight on backreaction effects by treating the semiclassical equations as a modified gravity theory motivated by the physics of the quantum vacuum. In this sense, analytic RSET approximations, along with the associated semiclassical equations, prove to be an invaluable tool.

In this work, we hence approximate the full four dimensional RSET through the prescription introduced in~\cite{Riegert1984} and further developed in~\cite{BalbinotFabbriShapiro1999,MottolaVaulin2006,AndersonMottolaVaulin2007}. This approximation is based on the choice of a --- non-local --- effective action that reproduces the trace anomaly (see Section~\ref{Sec:Preliminaries}), and on the introduction of auxiliary fields\footnote{The term ``auxiliary" is usually reserved for fields without dynamics. In this case, however, the auxiliary fields do propagate.} to cast the action in a local form. Varying this anomaly-induced effective action (AIEA) with respect to the metric yields a stress-energy tensor (SET) --- the anomalous stress-energy tensor (ASET) in what follows --- that depends on the auxiliary fields. When these fields are taken on-shell of their field equations, the trace of the ASET completely reproduces the trace anomaly. In this way, an analytical approximation for the exact RSET is obtained without the need to perform onerous sums and integrals over modes. 

Selecting a unique solution for the auxiliary fields requires specifying their boundary conditions. This choice is expected to encode properties of the vacuum states~\cite{BalbinotFabbriShapiro1999,ShenIzumiChen2015}. In a rough sense, if the amount of conditions that need to be imposed on the ASET to reproduce the same asymptotics as the exact RSET does not exceed the number of integration constants in the fields, then the AIEA approximation will succeed in yielding an ASET that reproduces the exact, numerically determined, RSET (to some degree). In fact, this approximation can be successfully applied to Schwarzschild spacetimes~\cite{BalbinotFabbriShapiro1999} to generate ASETs that agree, at leading order, with the exact RSET~\cite{Candelas1980} of conformally invariant scalar fields. 

Let us stress that, although in two dimensions there exists a direct correspondence between vacuum states and the auxiliary field~\cite{Barcelo:2011bb}, in four dimensions there is no direct reason why this should be expected~\cite{Bardeen:2018gca}, since
the ASET is specified up to traceless contributions.
In other words, given this ambiguity, we end up exploring a landscape of modified gravity theories, which, in simple spacetimes like Schwarzschild, properly reproduce the asymptotic properties of exact RSETs. However, we will see that this is not always guaranteed in more complex spacetimes like Reissner-Nordstr{\"o}m. 

It is worth mentioning that the AIEA approximation is so far the only truly four-dimensional scheme that allows to describe the Boulware, Hartle-Hawking and Unruh vacuum under the same prescription, and also the only one that is, in principle, applicable to black hole interiors. 

Summarizing our findings for each vacuum state:
\begin{itemize}
    \item Hartle-Hawking state: The AIEA approximation is perfectly regular at the event horizon and thus can be extended through it. We obtained an ASET that reproduces the Hartle-Hawking thermal bath at infinity and exhibits a divergent behaviour at the Cauchy horizon, both qualitatively and quantitatively consistent with the result obtained for minimally coupled scalar fields~\cite{Zilbermanetal2022}. Notice that previous approximations which describe this state, like the  Page~\cite{Page1993} and Anderson-Hiscock-Samuel~\cite{Andersonetal1995} ones, do not have such a straightforward extension beyond the event horizon, with the latter even predicting an unphysical logarithmic divergence there. 
    \item Unruh state: We show that the AIEA approximation, contrarily to previous claims in the literature~\cite{AndersonMottolaVaulin2007}, is incapable of reproducing the asymptotic properties of the Unruh state at Cauchy horizons and at infinity simultaneously. Therefore, we deem it as an inadequate approximation in Reissner-Nordstr\"om black holes. For this state, only the Polyakov approximation is known to succeed in predicting the correct $s$-wave contribution to the Hawking flux at infinity when backscattering effects are neglected. At the Cauchy horizon, instead, it predicts negative constant fluxes which appear to disagree with the positive fluxes obtained for minimally coupled fields~\cite{Zilbermanetal2022}. 
\end{itemize}

Approximations based on AIEA have been claimed to offer a consistent method to approximate the RSET in arbitrary spacetimes as long as the wave equations for the auxiliary fields can be integrated~\cite{MottolaVaulin2006}. Despite the clear advantages of the method, we show that the characteristics of the background (in particular, the presence of Cauchy horizons) plays an active role in its applicability, which casts doubts on its validity in more complex situations like dynamical or axisymmetric spacetimes. 

The paper is organized as follows. Section~\ref{Sec:Preliminaries} introduces the AIEA prescription and includes a subsection about the Reissner-Nordstr\"om spacetime. In Section~\ref{Sec:1+1}, we first apply the method to $1+1$ dimensions to show its equivalence with known results. In Subsection~\ref{Subsec:3+1}, we describe the application ---- and the limitations ---- of the method in $3+1$ dimensions. In~\ref{Subsec:Page}, we address the connection of this method with Page's approximation. In Subsections~\ref{Subsec:Aux} and~\ref{Subsec:HHStates}, we apply the method in $3+1$ dimensions to obtain the ASET in the Hartle-Hawking state. In~\ref{Subsec:UnruhStates}, we show that the method cannot give a suitable approximation of the RSET in the Unruh state. We conclude in Section~\ref{Sec:Conclusions} with a discussion on the results and some considerations.

\section{Preliminaries}
\label{Sec:Preliminaries}
In this section, we first explain the rationale behind the AIEA method. Then, for convenience, we briefly discuss Reissner-Nordstr{\"o}m spacetime and the conditions required to have a regular RSET at horizons.

\subsection{The anomalous effective action method}
\label{Subsec:Method}

As mentioned in the introduction, calculating the RSET is typically a hard task which requires significant computational power. The approximation based on the AIEA avoids these difficulties by relying on the fact that the RSET can be equally obtained from the quantum effective action $\Gamma$ by
\begin{equation}
    \langle\hat T^{\mu\nu}\rangle=\frac{2}{\sqrt{-g}}\fdv{\Gamma}{g_{\mu\nu}},
\end{equation}
Even assuming that one can make rigorous sense of it from a mathematical point of view, evaluating the full effective action would still be a daunting task. Nevertheless, we adopt the effective field theory (EFT) point of view and assume that $\Gamma[g]$ admits a derivative expansion in the metric field. General covariance fixes the form of the allowed terms at each order in the derivatives so that we are only left with the problem of determining the coefficients. What went under the carpet is that the expansion actually contains an infinite number of terms.
However, as long as we consider low energy physics, we can truncate this series in a meaningful way because terms with more derivatives become less relevant at low energies. 

This EFT construction assumes that high-energy physics decouples from low-energy physics, which is not the case in the presence of \textit{anomalies}. These are quantum effects, typically associated to symmetry breaking, which are strongly suppressed from a phenomenological perspective but become noticeable when the classical contribution is absent~\cite{Adler:1969gk, Wess:1971yu, Witten:1983tw, Bardeen:1984pm, Donoghue:1994Dn}.

In this work, we focus solely on the \textit{trace anomaly}~\cite{Duff1993}, which occurs when (classically) Weyl invariant matter fields are quantized on a curved background. In this case, the resulting RSET fails to be covariantly conserved and traceless at the same time. Since conservation follows from the principle of general covariance, which is expected to hold at low energies~\cite{Will:2014kxa}, it is more natural to relax Weyl invariance, allowing a non-zero trace of the RSET
\begin{equation}
\label{Eq:AnomalyD}
    \langle\hat T^\mu_{~\mu}\rangle=\mathcal A_d.
\end{equation}
The trace anomaly $\mathcal A_d$ depends on the dimension, but it is generically a local polynomial in the curvature tensor.
Despite being an ultraviolet phenomenon (the corresponding terms are \textit{marginal} in the EFT parlance), the anomaly plays a role that grows logarithmically with the length scale~\cite{BrownOttewill1983}, which means that it is \textit{marginally relevant} at low energies and must be taken into account in an effective description. As we mentioned, it is possible that, despite its weakness, the anomaly dominates over other classical effects. This is possible because of its intrinsic \textit{non-locality}: near horizons, where the curvature is small, classical effects are expected to be weak, but the causal structure differs significantly from that of flat space, thereby allowing for notable quantum effects.

To illustrate the rationale of the approximation, let us consider the following identity
\begin{equation}
\label{Eq:AnomalyWZrelation}
    \sqrt{-g}\langle\hat T^\mu_{~\mu}\rangle[g]=2g_{\mu\nu}\fdv{\Gamma[g]}{g_{\mu\nu}}=\fdv{\Gamma[e^{2\sigma}g]}{\sigma}\eval_{\sigma=0}.
\end{equation}
By combining this with Eq.~\eqref{Eq:AnomalyD}, we can formally integrate the anomaly polynomial along a Weyl transformation ($g\to g^\sigma\equiv e^{2\sigma}g$) to obtain the difference of the quantum effective action between the two conformally related metric, known as the Wess-Zumino (WZ) action
\begin{equation}
\label{Eq:WZDefinition}
    \int^{-\omega}_0\dd\sigma \sqrt{-g^\sigma}\langle\hat T^\mu_{~\mu}\rangle[g^\sigma]=\Gamma[g]-\Gamma[e^{-2\omega}g]\equiv\Gamma_{WZ}[g;\omega].
\end{equation}
Since the anomaly is a known polynomial in the curvature, it follows that the Wess-Zumino action is a known \textit{universal} term, in the sense that is does not depend on the vacuum state. In order to be more explicit, we need to consider the form of the anomaly polynomial on a case-by-case basis, which we will do in the following. For the remainder of this section, we will continue describing the method in a generic setting.

The Wess-Zumino effective action satisfies the following relation
\begin{equation}
\label{Eq:WZConsistencyCondition}
    \Gamma_{\textsc{wz}}[e^{-2\omega_2}g,\omega_1]-\Gamma_{\textsc{wz}}[g,\omega_1]=\Gamma_{\textsc{wz}}[e^{-2\omega_1}g,\omega_2]-\Gamma_{\textsc{wz}}[g,\omega_2]
\end{equation}
which expresses the closure of $\Gamma_{WZ}$ as a form in the cohomology of the local Weyl group. However, the Wess-Zumino action is not exact, in the sense that it cannot be written as the Weyl variation of any local\footnote{A non-local functional with this property obviously exists: the full effective action $\Gamma$.} functional. As a corollary, this implies that the anomaly $\mathcal A_d$ is a truly quantum effect, which cannot be entirely removed by the inclusion of (finite) local counterterms in the effective action.

Let us thus take $S_{\mathrm{nonloc}}$ to be one of the non-local functionals satisfying $\Gamma_{\textsc{wz}}[g;\omega]=S_{\mathrm{nonloc}}[g]-S_{\mathrm{nonloc}}[e^{-2\omega}g]$. By comparing with the definition~\eqref{Eq:WZDefinition}, it becomes clear that, for Weyl invariant matter fields
\begin{equation}
\label{Eq:splitting}
    \Gamma[g]=S_{\mathrm{nonloc}}[g]+S_{\mathrm{inv}}[g],
\end{equation}
where $S_{\mathrm{inv}}$ is a Weyl invariant functional of the metric. By construction, the separation between $S_{\mathrm{nonloc}}$ and $S_{\mathrm{inv}}$ is ambiguous, as Weyl invariant part can be removed from the latter and added to the former, but we assume that a specific choice has been made. After that, one can cast $S_{\mathrm{nonloc}}$ in a local form by the introduction of a set of auxiliary fields $\{\phi\}$~\cite{FradkinTseytlin1983,Riegert1984,BalbinotFabbriShapiro1999}. In more details, one construct a local functional $S_{\mathrm{anom}}[g;\{\phi\}]$ such that, when evaluated on-shell of the auxiliary fields, it reproduces the original non-local functional $S_{\mathrm{nonloc}}[g]$.
The AIEA approximation then consists in neglecting the contribution from $S_{inv}$ to the full effective action. In other words, we declare that the RSET is approximated by the ASET
\begin{equation}
    \langle \hat T^{\mu\nu} \rangle\approx T_{\mathrm{anom}}^{\mu\nu},\quad \text{where}\quad T_{\mathrm{anom}}^{\mu\nu}=\frac{2}{\sqrt{-g}}\fdv{S_{\mathrm{anom}}[g;\{\phi\}]}{g_{\mu\nu}}\eval_{\mathrm{on-shell}\;\{\phi\}}.
\end{equation}

We remark that the choice of splitting in Eq.~\eqref{Eq:splitting} is arbitrary;
the only justification for it will stem from the consistency of the predictions it makes with previous works~\cite{BalbinotFabbriShapiro1999} and from the agreement with exact results. In this sense, the AIEA prescription can be seen as a set of approximations, each associated with a choice of $S_{\mathrm{anom}}$ and $S_{\mathrm{inv}}$. Among these, only some will produce ASETs that adequately approximate the full RSET.

Different formulations of this approximation were used and analysed extensively in~\cite{BalbinotFabbriShapiro1999,Balbinot:1999vg, MazurMottola2001, MottolaVaulin2006, AndersonMottolaVaulin2007, Anderson:2007te}, and were later applied to cosmology \cite{Mottola:2010qg} and black hole collapse \cite{Mottola:2023jlo}. In Sections~\ref{Sec:1+1} and~\ref{Sec:3+1}, we will summarise the basics of the AIEA prescription in $1+1$ and $3+1$ dimensions. Now, we take a brief detour to introduce the spacetime to which the method will be applied.

\subsection{The Reissner-Nordstr{\"o}m spacetime}

\begin{figure}
    \centering
    \includegraphics[width=0.8\linewidth]{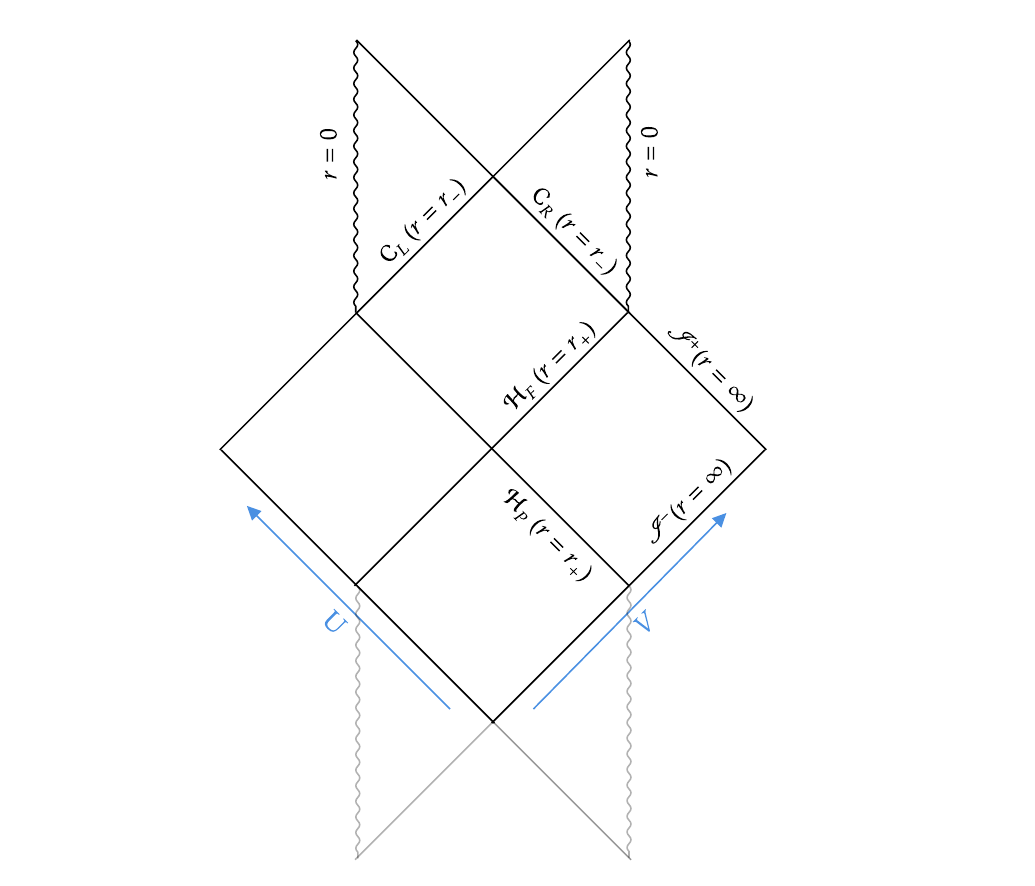}
    \caption{Penrose diagram of the maximal analytical extension of the Reissner-Nordstr\"om spacetime. The future ($U=0$) and past ($V=0$) branches of the event horizon $r_+$ are denoted by $\mathcal H_F$ and $\mathcal H_P$. The two branches of the Cauchy horizon are denoted by $\mathcal C_R$ and $\mathcal C_L$, where the subscripts stand respectively for \textit{right} and \textit{left} wedges.}
    \label{Fig:RNspacetime}
\end{figure}

We consider spherically symmetric spacetimes of the form
\begin{equation}
    ds^{2}=-f(r)dt^{2}+h(r)dr^{2}+r^{2}d\Omega^{2},
\end{equation}
where $d\Omega^{2}$ is the angular line element the 2-sphere. 
The Reisnser-Nordstr{\"o}m (RN) spacetime is the only static and spherically symmetric solution to the Einstein-Maxwell equations that is asymptotically flat. Given its simplicity, it is an optimal setting to investigate quantum vacuum effects in the proximity of Cauchy horizons without renouncing to the benefits of spherical symmetry. The metric functions of RN spacetime are
\begin{equation}\label{Eq:SphSymMetric}
    f(r)=h(r)^{-1}=\frac{\left(r-r_{+}\right)\left(r-r_{-}\right)}{r^{2}},
\end{equation}
where 
\begin{equation}
        r_{\pm} = M \pm \sqrt{M^2-Q^2},
\end{equation}
denote the positions of the outer (upper sign) and inner (lower sign) horizons.
For a static spacetime, the outer horizon is an event horizon, an the inner horizon is a Cauchy one.\footnote{Throughout this paper, we will use the terms event and Cauchy to refer to the outer and inner future trapping horizons of the Reissner-Nordstr{\"o}m spacetime. When necessary, we will refer to the right or left wedges of the Cauchy horizon explicitly.} The surface gravity at these horizons is given by
\begin{equation}
    \kappa_{\pm} = \frac{r_{\pm}-r_{\mp}}{2r_{\pm}^{2}}.
\end{equation}
Upon introducing the tortoise coordinate
\begin{equation}
   \label{Eq:Tortoise} 
   r^* = r + \frac{1}{2\kappa_+}\log\abs{\frac{r}{r_+} - 1} - \frac{1}{2\kappa_-}\log\abs{\frac{r}{r_-} - 1},
    \end{equation}
one can write the outgoing and ingoing Eddington-Finkelstein null coordinates as
\begin{equation}
    u=t-r^{*},\quad v=t+r^{*}. 
\end{equation}
The RN metric is manifestly regular at the event horizon when expressed in terms of horizon-penetrating Kruskal coordinates
\begin{equation}
        U_+ = -\kappa_+^{-1}e^{-\kappa_+ u}, \quad V_+ = \kappa_+^{-1} e^{\kappa_+ v},
        \label{Eq:Kruskal+}
\end{equation}
which cover the region $r_{-}<r<\infty$ in Fig.~\ref{Fig:RNspacetime}. A second set of Kruskal coordinates, covering the region $0<r<r_{+}$, is
\begin{equation}
        U_- = -\kappa_-^{-1} e^{\kappa_- u}, \quad V_- = -\kappa_-^{-1} e^{-\kappa_- v}.
\end{equation}
Neither of these Kruskal coordinates fully cover the maximally extended Reissner-Nordstr{\"o}m metric, but they can be used locally to determine whether physical quantities, such as the RSET, are regular in the reference frame of freely falling observers crossing the horizons.

The regions we are interested in are the future event horizon $\mathcal H_F$ and the left and right wedges of the Cauchy horizon $\mathcal C_{L,R}$ (see Fig.~\ref{Fig:RNspacetime}). The regularity conditions at $\mathcal H_F$ are
\begin{equation}\label{Eq:Regularity+}
\langle\hat{T}_{uu}\rangle\sim\left(r-r_{+}\right)^{2},\quad \langle\hat{T}_{uv}\rangle\sim\left(r-r_{+}\right)^{1},\quad \langle\hat{T}_{vv}\rangle\sim\left(r-r_{+}\right)^{0},
\end{equation}
while, at the left wedge $\mathcal C_L$ of the Cauchy horizon, they are
\begin{equation}\label{Eq:Regularity-}
\langle\hat{T}_{uu}\rangle\sim\left(r-r_{-}\right)^{0},\quad \langle\hat{T}_{uv}\rangle\sim\left(r-r_{-}\right)^{1},\quad \langle\hat{T}_{vv}\rangle\sim\left(r-r_{-}\right)^{2}.
\end{equation}
The conditions for regularity at $\mathcal C_R$ and $\mathcal H_P$ (the past event horizon) are identical to~\eqref{Eq:Regularity+} and~\eqref{Eq:Regularity-}, respectively, after interchanging $r_{+}$ and $r_{-}$.

\section{The anomaly-induced effective action in 1+1 dimensions}
\label{Sec:1+1}

One strategy that has been frequently used to study spherically symmetric (black hole) spacetimes is dimensional reduction~\cite{Balbinot:2000iy,Frolov:1999an,Sutton:2000gm}. In this strategy, one considers the $(t,r)$ sector of the metric~\eqref{Eq:SphSymMetric} as a $1+1$ dimensional manifold, i.e.,
\begin{equation}\label{Eq:ConfFlatMetric}
    \dd s^2=\dd s^{2}_{(2)}+r^{2}d\Omega^{2},
\end{equation}
where, in double-null Eddington-Finkelstein coordinates,
\begin{equation}\label{Eq:2DMetric}
     \dd s^{2}_{(2)}=-C(u,v)\dd u\dd v.
\end{equation}
Ignoring the contribution from non spherical waves and the scattering from the gravitational potential, one can approximate the full $3+1$ RSET by the so-called Polyakov RSET~\cite{Polyakov1981,FabbriNavarro-Salas2005}
\begin{equation}\label{Eq:Polyakov}
    {T^{\mu}_{~\nu}}^{(\rm P)}\equiv \frac{\delta_{a}^{~\mu}\delta_{\nu}^{~b}\langle\hat{T}^{a}_{~b}\rangle^{(2)}}{4\pi r^{2}},
\end{equation}
where $\langle\hat{T}^{\mu}_{~\nu}\rangle^{(2)}$ is the RSET evaluated in the $2$-dimensional spacetime~\eqref{Eq:ConfFlatMetric}, and we included projectors from the indices in this spacetime (latin letters) to the indices in the $3+1$ spacetime (greek letters). The factor $4\pi r^{2}$ is introduced to ensure the conservation of ${T^{\mu}_{~\nu}}^{(\rm P)}$ at the cost of making the RSET singular at $r=0$.\footnote{In static spacetimes we can cutoff this factor~\cite{Arrecheaetal2022} to regularize the RSET, while re-introducing at the same time some angular components.} The Polyakov approximation~\eqref{Eq:Polyakov} (also known as $s$-wave approximation) is particularly good near horizons, where the scattering effects of the gravitational potential are negligible and the wave equation for scalar fields reduces to the two-dimensional one~\cite{FabbriNavarro-Salas2005}. As a consequence, it properly captures the regular behaviour of the Hartle-Hawking and Unruh states (see~\ref{Subsec:States}) and has been used to numerically simulate Hawking evaporation in four dimensions~\cite{ParentaniPiran1994}.
Its divergence at $r=0$ plays no role in the Reissner-Nordstr{\"o}m spacetime since the states we consider are already singular at the Cauchy horizon.

In this Section, we will first show how the exact $2$D RSET can be obtained from the AIEA prescription. Although part of the material already exists in the literature, we find it pedagogical to the discussion in the $3+1$ dimensional case. The initial part of the analysis applies to arbitrary spacetimes. In the final part, we specialize to RN and determine the form of the RSET divergence at the Cauchy horizon.

\subsection{Prescription in 1+1 dimensions}
\label{Subsec:1+1}
In $1+1$ dimensions, the trace anomaly is given by
\begin{equation}
\label{Eq:TraceAnomin4Dalyin2D}
    \langle\hat T^{\mu}_{~\mu}\rangle=\mathcal A_2=\frac{N}{24\pi} R,
\end{equation}
where $N=N_s+N_f$, and $N_{s/f}$ is the number of massless scalars/fermions in the theory. Since there is no local, generally covariant action whose variation can produce the Ricci scalar, this anomaly describes a genuinely non-local quantum effect. The WZ action can be obtained integrating~\eqref{Eq:TraceAnomin4Dalyin2D} using the relation
\begin{equation}
\label{Eq:RicciScalarConformal}
    \sqrt{-g^\sigma}R[g^\sigma]=\sqrt{-g}(R[g]-2\square\sigma),
\end{equation}
where we used again the shortcut $g^\sigma\equiv e^{2\sigma}g$.
Quite trivially, this leads to
\begin{equation}
    \Gamma_{\textsc{wz}}[g;\omega]=\frac{N}{24\pi}\int\dd[2]x\sqrt{-g}(\omega R+\omega\Box\omega)
\end{equation}
While it is straightforward to check that this functional satisfies the Wess-Zumino consistency condition~\eqref{Eq:WZConsistencyCondition}, a much less trivial task is to find the non-local action $S_{\mathrm{nonloc}}$. Luckily, in two dimensions, a suitable expression can be obtained by inverting \eqref{Eq:RicciScalarConformal} and using the conformal invariance of $\sqrt{-g}\,\square$ in two (and only two) dimensions. The result is the celebrated Polyakov action~\cite{Dowker1993}
\begin{equation}
\label{Eq:NonlocalAction2D}
    S_{\mathrm{nonloc}}[g]=-\frac{N}{96\pi}\int\dd[2]x \sqrt{-g(x)}\int\dd[2]x' \sqrt{-g(x')} R(x)\square^{-1}(x,x')R(x').
\end{equation}
Here $\square^{-1}(x,x')$ denotes the Green function that inverts the differential operator $\square$.\footnote{Explicitly
\begin{equation}
    \int\dd[2]x\sqrt{-g(x)}\,\square^{-1}(x,x')\square f(x)=f(x').
\end{equation}}
To cast $S_{\mathrm{nonloc}}$ in a local form, we need to introduce an auxiliary field satisfying 
\begin{equation}\label{Eq:Field2D}
    \square\varphi=-R.
\end{equation}

The AIEA that implements this field equation is
\begin{equation}\label{Eq:Action2D}
    S_{\mathrm{anom}}[g;\varphi]=-\frac{N}{96\pi}\int\dd[2]x \sqrt{-g}(\nabla^\mu\varphi\nabla_\mu\varphi-2R\varphi).
\end{equation}
One can check that, by replacing the auxiliary field with the general solution of~\eqref{Eq:Field2D}, the anomalous action reproduces \eqref{Eq:NonlocalAction2D} up to a boundary term. 
Varying the action~\eqref{Eq:Action2D} with respect to the metric yields the ASET
\begin{equation}
\label{Eq:ASETin2D}
    T^{\mathrm{anom}}_{\mu\nu}=\frac{N}{24\pi}\pqty{\nabla_\mu\nabla_\nu\varphi-g_{\mu\nu}\Box\varphi+\frac{1}{2}\nabla_\mu\varphi\nabla_\nu\varphi-\frac{1}{4}g_{\mu\nu}\nabla^\lambda\varphi\nabla_\lambda\varphi},
\end{equation}
whose trace completely reproduces the quantum anomaly when $\varphi$ is on-shell:
\begin{equation}
    g^{\mu\nu}T^{\mathrm{anom}}_{\mu\nu}=-\frac{N}{24\pi}\Box\varphi=\frac{N}{24\pi}R.
\end{equation}


It is worth noting that, given the field equation for $\varphi$, the metric $e^{-\varphi}g$ has the property of making the anomaly vanish when the field is on-shell:\footnote{This relation is the same that underlies Page's approximation, as noted in Eq.~(2.13) of Ref.~\cite{BrownOttewill1985}}
\begin{equation}
\label{Eq:PageCondition2D}
    \mathcal A_2[e^{-\varphi}g]=0.
\end{equation}
Thus, $\varphi$ can be seen as an order parameter that provides a physical (i.e. generally covariant) parameterization for the conformal class of $g$. 
Since the auxiliary field solutions cannot be expressed in terms of local geometrical invariants, the low-energy effective description containing the anomalous part can describe long-range effects that we understand as arising from the quantum, yet macroscopic, coherence of the matter state. For example, the dependence on $\varphi$ on the massless propagator $\square^{-1}$ --- which diverges on the light cone --- can generate significant effects at horizons. These would contradict the naive intuition that quantum effects are sizeable \textit{only} in regions where the curvature becomes Planckian.

\subsection{The ASET components}
In $1+1$ dimensions, we can leverage on the fact that every metric is conformally flat to find the exact analytical solution to the auxiliary field equation~\eqref{Eq:Field2D}. Namely, using double null-coordinates, we can write
\begin{equation}
\label{Eq:Conformalon2Dsolution}
    \varphi=\ln C(u,v)+A(u)+B(v),
\end{equation}
where $A$ and $B$ are generic functions that correspond to homogeneous solutions of the field equation and $C$ is the conformal factor in~\eqref{Eq:2DMetric}. The form of $A$ and $B$ is determined by the boundary conditions we fix to invert the $\square$ operator. A straightforward computation then leads to the following ASET components
\begin{subequations}
\label{Eq:ASETin2Duuvv}
\begin{align}
    &T^{\mathrm{anom}}_{uu}=\frac{N}{24\pi}\pqty{\frac{\partial^2_u C}{C}-\frac{3}{2}\pqty{\frac{\partial_u C}{C}}^2+\partial_u^2 A+\frac{1}{2}\pqty{\partial_u A}^2},\\
    &T^{\mathrm{anom}}_{vv}=\frac{N}{24\pi}\pqty{\frac{\partial^2_v C}{C}-\frac{3}{2}\pqty{\frac{\partial_v C}{C}}^2+\partial_v^2 B+\frac{1}{2}\pqty{\partial_v B }^2},\\
    &T^{\mathrm{anom}}_{uv}=-\frac{N}{24\pi}\pqty{\frac{\partial_u \partial_v C}{C}-\frac{\partial_u C}{C}\frac{\partial_v C}{C}}.
\end{align}
\end{subequations}
The $A$ and $B$ dependent terms can also be written as the Schwarzian derivatives $\left\{U,u\right\}$ and $\left\{V,v\right\}$ 
between the Eddington-Finkelstein coordinates $(u,v)$ and the Kruskal-like coordinates $(U,V)$, which are defined through the relations
\begin{equation}
\label{Eq:KruskalConformalCoordinates}
    \dd U=e^{-A(u)}\dd u,\quad \dd V=e^{-B(v)}\dd v.
\end{equation}
Indeed
\begin{equation}
    \partial_u^2 A+\frac{1}{2}\pqty{\partial_u A}^2=\frac{\dd^{3}U}{du^{3}}\bigg/\frac{\dd U}{du}-\frac{3}{2}\left(\frac{\dd^{2}U}{du^{2}}\bigg/\frac{\dd U}{du}\right)^{2}=\left\{U,u\right\},
\end{equation}
and analogously for $V(v)$. 
This makes the two-dimensional ASET entirely equivalent to the RSET obtained through point-splitting regularization~\cite{DaviesFulling1977,FabbriNavarro-Salas2005}. This equivalence is expected since we know that the $1+1$-dimensional RSET is fixed by covariant conservation and the trace relation~\eqref{Eq:TraceAnomin4Dalyin2D}, up to state-dependent flux terms in the $uu, vv$ components.

Even if the AIEA prescription makes no reference to quantum states, as it starts from an educated guess of the infrared effective action, we conclude --- in light of expressions~\eqref{Eq:Conformalon2Dsolution} and~\eqref{Eq:ASETin2Duuvv} --- that the method encodes a notion of state dependence via the boundary conditions required to determine a unique solution to the auxiliary field equation. The authors of~\cite{ShenIzumiChen2015} give another interesting approach to realize this connection.

\subsection{Vacuum states in the AIEA prescription}
\label{Subsec:States}

Let us particularize this discussion to static black hole spacetimes described by the line element
\begin{equation}
    \dd s^{2}_{(2)}=-f(r)dt^{2}+f(r)^{-1}dr^{2}.
\end{equation}
We require that the ASET does not depend explicitly on time for consistency with the symmetry of the background. If so, the only allowed time dependence in $\varphi$ is linear, which implies that both $A(u)$ and $B(v)$ in \eqref{Eq:Conformalon2Dsolution} must be linear functions of their arguments. For convenience, we write them as
\begin{equation}\label{Eq:ABSols2D}
    A=\frac{1}{2}\left(q-p\right)u,\quad B=\frac{1}{2}\left(q+p\right)v,
\end{equation}
so that the field solution reads (we set an irrelevant constant to zero)
\begin{equation}
\label{Eq:Conformalon2DsolutionLinear}
    \varphi=\ln f+q r^*+pt.
\end{equation}
Replacing solution~\eqref{Eq:Conformalon2DsolutionLinear} in~\eqref{Eq:ASETin2D}, we obtain the ASET in double null coordinates $(u,v)$\footnote{From now on, we will omit writing the subscript `anom' since the stress-energy tensor under consideration will always be the ASET.}
\begin{equation}
\label{Eq:1+1ASET}
    T_{\mu\nu}=\frac{N}{192\pi}\mqty(2ff''-{f'}^2+\left(q-p\right)^{2} & 2ff''\\ 2ff'' & 2ff''-{f'}^2+\left(q+p\right)^{2}),
\end{equation}
where the $'$ denotes a derivative with respect to the radial coordinate $r$.

We are interested in the cases where $f$ has one or more roots, coinciding with the Killing horizons of the Killing vector field $\partial_{t}=\partial_{u}+\partial_{v}$ that generates translations in $t$. Assuming the following near-horizon expansion for the conformal factor $f$
\begin{equation}
    f=2\kappa_{\textsc{h}} \left(r-r_{\textsc{h}} \right)+\order{r-r_{\textsc{h}}^{2}},
\end{equation}
the ASET will satisfy the regularity conditions analogous to~\eqref{Eq:Regularity+}
as long as
\begin{equation}\label{Eq:RegCond2DH}
    4\kappa_{\textsc{h}}^{2}-\left(q\mp p\right)^{2}=0,
\end{equation}
where the upper (lower) sign is needed for regularity at the future (past) branch of the event horizon. 

Since $p$ and $q$ are constants, it is clear that these solutions cannot be regular at multiple horizons ($r_{{\textsc{h}},1}\ne r_{{\textsc{h}},2}$) unless they share the same surface gravity ($\kappa_{{\textsc{h}},1}=\kappa_{{\textsc{h}},2}$). In view of~\eqref{Eq:RegCond2DH}, there exist infinite choices of $\{p,q\}$ values which yield singular ASETs at the event horizon, corresponding to thermal states that are not in equilibrium with the horizon temperature~\cite{Loranz:1995gc,Bazarov:2021rrb}.\footnote{Notice that the Boulware state is just the zero temperature case within this broader family of states.} At the asymptotically flat region, where the conformal factor is approximated by $f= 1+\order{r^{-1}}$, the conditions
\begin{equation}\label{Eq:BCond2D}
    \text{B}: \{p=q=0\}
\end{equation}
guarantee the ASET behaves asymptotically as
\begin{equation}
    T_{uu}=T_{vv}=T_{uv}=\order{r^{-3}}.
\end{equation}
This reproduces the behaviour of the Boulware state, which is indeed singular at both the past and future event horizons. The only vacuum that is simultaneously regular on the two branches is the Hartle-Hawking state, which we can reproduce with the conditions
\begin{equation}\label{Eq:HHCond2D}
    \text{HH}: \{q=\mp 2 \kappa_{\textsc{h}},~p=0\}\quad \text{or} \quad \{q=0,~p=\mp 2\kappa_{\textsc{h}}\}.
\end{equation}
In both cases, the outflux of the solution vanishes
\begin{equation}
    T_{rt}=T_{vv}-T_{uu}=\frac{N}{48\pi}pq=0.
\end{equation}
Conversely, when such a flux is non-vanishing, the state must become singular at one of the branches of the event horizon. This is what happens in the Unruh state, in which one also requires that $T_{uu}$ vanishes at infinity. This state is obtained by the choice
\begin{equation}\label{Eq:UCond2D}
    \text{U}:\{p=-q=\pm\kappa_{\textsc{h}}\}.
\end{equation}

We have seen that the AIEA prescription generates an ASET that is unique up to a collection of integration constants in the auxiliary field~\eqref{Eq:ASETin2Duuvv}. In order to retrieve the RSET corresponding to the three most studied vacuum states, we need to impose conditions on the ASET either at future and past infinity for Boulware, at both branches of the event horizon for Hartle-Hawking, or at the future event horizon and past infinity for Unruh. In a way, this is equivalent to selecting the global behaviour of the auxiliary field as to reproduce the correct asymptotic properties of the desired vacuum state, which are known \textit{beforehand} through canonical quantisation analyses. In $3+1$ dimensions, as we will see, the situation is considerably more complex. To select the ``true" Hartle-Hawking or Unruh states, we will need to impose conditions at the Cauchy horizon as well. The underlying reason is that, in $3+1$ dimensions, the RSET has a larger variety of possible divergent behaviours at horizons than in $1+1$, thus requiring more conditions on the ASET to match the expected result.

Before continuing our analysis of the Hartle-Hawking and Unruh states, let us observe that the conditions in \eqref{Eq:RegCond2DH} have a nice geometrical interpretation in terms of the conformally transformed metric $e^{-\varphi}g$.
This conformally related spacetime is flat almost everywhere, but, since it is not physical, it is not required to be regular. We briefly show how the conditions \eqref{Eq:RegCond2DH} on the ASET are equivalent to the absence of defects in the Euclidean section of the conformally-related metric: using Kruskal-like coordinates $U,V$ we can write
\begin{equation}
\label{Eq:ConformalLineElement2D}
    e^{-\varphi}\dd s^2=-(-\kappa_{\textsc{h}} U)^{-\alpha} (\kappa_{\textsc{h}} V)^{-\beta} \dd U\dd V,\quad\text{with}\quad \alpha=\frac{q-p}{2\kappa_{\textsc{h}}}+1,\;\beta=\frac{p+q}{2\kappa_{\textsc{h}}}+1.
\end{equation}
Assuming that $\alpha,\beta\ne 1$ (i.e. $p^2\ne q^2$), we introduce a new set of coordinates, defined by
\begin{equation}
\label{Eq:Flattenize}
    \kappa_{\textsc{h}} U=-\rho^{\frac{1}{1-\alpha}}e^{\frac{1}{1-\alpha}\tau},\qquad \kappa_{\textsc{h}} V=\rho^{\frac{1}{1-\beta}}e^{-\frac{1}{1-\beta}\tau},
\end{equation}
which makes spacetime manifestly flat: the line element becomes $\dd\rho^2-\rho^2\dd\tau^2$ up to a constant. However, the coordinate transformation~\eqref{Eq:Flattenize} fails to extend to a regular Euclidean section unless 
$\alpha,\beta=0$ or $\alpha,\beta=2$. In terms of $p$ and $q$, these cases correspond to a no-flux situation. For example, let us consider the cases when $\alpha=\beta$, corresponding to the first Hartle-Hawking condition in~\eqref{Eq:HHCond2D}. Sending $\tau\to i\theta$, we see that $U$ and $V$ become complex conjugate coordinates and the orbits of $\partial_\theta$ are circles around the origin: the condition $\alpha=\beta$ ensures that $U$ and $V$ have the correct monodromy. The other cases (when $\alpha+\beta=2$) are obtained in a similar fashion after one of the Kruskal coordinate is inverted, say $U\to-1/U$.

\subsection{The Hartle-Hawking and Unruh stress-energy tensors at horizons}
Let us conclude this section by evaluating the ASET close to the two horizons of the Reissner-Nordstr{\"o}m metric~\eqref{Eq:SphSymMetric}.

Assuming the linear solution \eqref{Eq:Conformalon2DsolutionLinear}, the ASET components in $(u,v)$ coordinates close to the event horizon ($\kappa_{\textsc{h}}\to \kappa_+$) read
\begin{equation}
    T_{\mu\nu}\overset{r\to r_+}{\sim}\frac{N}{192\pi}\mqty((q-p)^2-4\kappa_+^2+\order{r-r_{+}}^{2} & \mathcal O\pqty{r-r_+}\\\mathcal O\pqty{r-r_+} & (q+p)^2-4\kappa_+^2+\order{r-r_{+}}^{2}).
\end{equation}
It is straightforward to see that the Boulware, Hartle-Hawking and Unruh states, defined by the conditions~\ref{Eq:BCond2D},~\ref{Eq:HHCond2D}, and~\ref{Eq:UCond2D} respectively, are characterized by the flux components
\begin{align}
    &T_{uu}^{\rm B}=
    T_{vv}^{\rm B}=T_{vv}^{\rm U}=-\frac{N\kappa_{+}^2}{48\pi},\\
    &T_{uu}^{\rm{HH}}=
    T_{vv}^{\rm {HH}}=T_{uu}^{\rm U}=\order{r-r_{+}}^{2}.
\end{align}

At the inner horizon we obtain analogous expansions, with $\kappa_{\textsc{h}}\to\kappa_-$:
\begin{equation}
    T_{\mu\nu}\overset{r\to r_-}{\sim}\frac{N}{192\pi}\mqty((q-p)^2-4\kappa_-^2+\order{r-r_{-}}^{2} & \mathcal O\pqty{r-r_-}\\\mathcal O\pqty{r-r_-} & (q+p)^2-4\kappa_-^2+\order{r-r_{-}}^{2}).
\end{equation}
For the Hartle-Hawking and Unruh states, we now obtain
\begin{align}\label{Eq:HHFluxes2D}
    T_{uu}^{\rm HH}=
    T_{vv}^{\rm HH}=T_{uu}^{\rm U}=\frac{N\left(\kappa_{+}^2-\kappa_{-}^{2}\right)}{48\pi},\quad T_{vv}^{\rm U}=-\frac{N\kappa_{-}^{2}}{48\pi}.
\end{align}
Due to the presence of fluxes, these states diverge on both wedges of the Cauchy horizon. Recall that the 
components of a 2D stress-energy can be mapped into those of a 4D one through Polyakov's approximation~\eqref{Eq:Polyakov}. Before extracting conclusions about the 4D RSET, a few observations are in order. First, since the above are local expansions around horizons, it should be possible to construct --- outside the context of vacuum GR --- a black hole spacetime with at least two horizons having the same surface gravity.
Doing this, one would obtain a Hartle-Hawking state that is regular at the Cauchy horizon as well, both on the left and right wedges.\footnote{Such a spacetime would still suffer from mass inflation instabilities as long as $\kappa_{-}\neq0$.} The Unruh state, on the other hand, would still be singular on the right wedge as long as $\kappa_-\ne 0$. Second, in 2D, the order (and the sign) of the divergence of the Boulware state at the event horizon is the same as that of the Unruh state at the Cauchy horizon --- in both cases, $T_{uu}$ approaches a (negative) constant. This equivalence is broken in 4D, where the RSET in the Unruh state, for which $T_{uu}$ and $T_{vv}$ are constant at the Cauchy horizon, turns out to have a milder singularity  than the one exhibited by the Boulware state, for which $T_{uu}$ and $T_{vv}$ are $\propto1/(r-r_{+})$ at the event horizon~\cite{Sela:2018xko}. Third, in $1+1$ dimensions, both Hartle-Hawking and Unruh states have flux components which are \textit{finite} and \textit{negative}. While the RSET of minimally and conformally coupled scalar fields in $3+1$ dimensions does exhibit finite fluxes at the Cauchy horizon~\cite{Sela:2018xko,Zilbermanetal2022,Klein:2023rwg}, the sign of these components is instead \textit{positive} for a wide range of $Q/M$ values. It is difficult to pinpoint the physical origin of this discrepancy, since field propagation is much more complex in $3+1$, involving backscattering effects and higher-multipole corrections. Moreover, the sign of the flux components might depend on the curvature coupling in a non-trivial way, which means that the previous observation should be taken with a grain of salt.

The $T_{vv}$ flux component is the main agent driving the backreaction of the geometry in the proximity of the Cauchy horizon: if positive, the horizon size will suddenly contract; if negative, it will expand~\cite{McMaken:2024fvq}. In dynamical collapse scenarios, as the \textit{in} state approaches the Unruh vacuum, inner horizons have been found to undergo an expanding phase~\cite{Boyanov:2022xfw,Barenboim:2024Dko} when sourced by the Polyakov SET, compatibly with a negative flux. Whether this effect will survive in a full $3+1$ analysis remains to be tested. 

\section{The anomaly-induced effective action in 3+1 dimensions}
\label{Sec:3+1}

In this section, we apply the AIEA prescription to $3+1$ dimensions. Unlike the 2D case, drawing a correspondence between boundary conditions for the auxiliary fields and vacuum states is more challenging in 4D. Our analysis reveals that, within the range of explored possibilities, the method can capture some features of the Hartle-Hawking state but seems unable to accurately describe the Unruh state.

\subsection{Prescription in 3+1 dimensions}
\label{Subsec:3+1}
In $3+1$ dimensions, the trace anomaly is given by \cite{Deseretal1976, Duff1993}
\begin{equation}
\label{Eq:TraceAnomalyin4D}
\langle\hat T^\mu_{~\mu}\rangle=\mathcal A_4=b_F F+b_E \pqty{E-\frac{2}{3}\square R}+b'\square R,
\end{equation}
where $F$ and $E$ are respectively by the squared Weyl tensor and the $4D$ Euler density:
\begin{subequations}
\label{Eq:EFdefinitions}
\begin{align}
F&=
C_{\mu\nu\rho\sigma}C^{\mu\nu\rho\sigma}=R_{\mu\nu\rho\sigma}R^{\mu\nu\rho\sigma}-2R_{\mu\nu}R^{\mu\nu}+\frac{R^2}{3},\\
E&={}^{*}R_{\mu\nu\rho\sigma}{}^{*}R^{\mu\nu\rho\sigma}=R_{\mu\nu\rho\sigma}R^{\mu\nu\rho\sigma}-4R_{\mu\nu}R^{\mu\nu}+R^2.  
\end{align}
\end{subequations}
The coefficients $b_F$ and $b_E$ in \eqref{Eq:TraceAnomalyin4D} are fixed by the number of massless, conformally coupled scalars, massless Dirac fermions, and massless vectors in the theory:
\begin{equation}
    b_F=\frac{1}{1920\pi^2}\left(N_{s}+6N_{f}+12N_{v}\right),\quad b_E=-\frac{1}{5760\pi^2}\left(N_{s}+11N_{f}+62N_{v}\right).
\end{equation}
Notice that $b_F\ge 0$, while $b_E\le 0$. On the other hand, the coefficient $b'$ is regularization-dependent because $\Box R$\,---\,unlike the other terms in~\eqref{Eq:TraceAnomalyin4D}\,---\,can be obtained as the variation of a local diffeomorphism-invariant quantity, namely $R^2$. This means that this term can be eliminated by adding a finite counterterm to the effective action. In the rest of the paper, we will simply set $b'=0$.

The steps that lead to the auxiliary field description of the non-local effective action are analogous to the 2D case, but more involved \cite{MottolaVaulin2006}.
Given the form of the anomaly~\eqref{Eq:TraceAnomalyin4D} with $b'=0$, we expect that a local description would require two auxiliary fields, respectively sourced by the two non-trivial cocycles of the local Weyl group \cite{MazurMottola2001}:
\begin{equation}
\label{Eq:Conformalon4Dequations}
    2\Delta_{4}\phi=\pqty{E-\frac{2}{3}\square R},\qquad 2\Delta_{4}\psi=F,
\end{equation}
where
\begin{equation}
    \Delta_{4}=\Box^2+2 R^{\mu\nu}\nabla_{\mu}\nabla_{\nu}-\frac{2}{3} R\Box+\frac{1}{3}\left(\nabla^{\mu}R\right)\nabla_{\mu}
\end{equation}
is the unique fourth order differential operator that is Weyl invariant in four (and only four) dimensions, i.e.  $\sqrt{-g^\sigma}(\Delta_4)_{g^\sigma}=\sqrt{-g}(\Delta_4)_g$. An appropriate local form of the anomalous action is
\begin{align}
\label{Eq:AnomAction4D}
    S_{\text{anom}}[g;\phi,\psi]
    &
    =
    \frac{b_E}{2}\int d^{4}x\sqrt{-g}\left\{-\phi\Delta_{4}\phi+\left(E-\frac{2}{3}\Box R\right)\phi\right\}\nonumber\\
    &
    +\frac{b_F}{2}\int d^{4}x\sqrt{-g}\left\{-2\phi\Delta_{4}\psi+F\phi+\left(E-\frac{2}{3}\Box R\right)\psi\right\}.
\end{align}
It is easy to see that this action enforces the field equations~\eqref{Eq:Conformalon4Dequations}. By varying with respect to the metric, we construct the ASET
\begin{equation}
    \label{Eq:EFtensor}
    T^{\mu\nu}_{\text{anom}}=\frac{2}{\sqrt{-g}}\frac{\delta S_{\text{anom}}}{\delta g_{\mu\nu}}=b_E E^{\mu\nu}+b_F F^{\mu\nu},
\end{equation}
where the tensors $E^{\mu\nu}$ and $F^{\mu\nu}$ are independently conserved. The reader can find the explicit expression of these tensors in~\cite{MottolaVaulin2006} and check that they reproduce the full trace anomaly on-shell of the auxiliary fields
\begin{equation}\label{Eq:EoMs4D}
    E^{\mu}_{~\mu}=2\Delta_{4}\phi\doteq E-\frac{2}{3}\square R,\qquad F^{\mu}_{~\mu}=2\Delta_{4}\psi\doteq F.
\end{equation}

The action \eqref{Eq:AnomAction4D} has been used in \cite{MottolaVaulin2006, AndersonMottolaVaulin2007}, but modifications of it were proposed already in \cite{BalbinotFabbriShapiro1999,Balbinot:1999vg} and~\cite{Bardeen:2018gca} to adsress some shortcomings in the predictions for the Unruh state in the Schwarzschild spacetime. Since then, critiques to the AIEA method have mostly focused on mismatches between the asymptotic contributions in the ASET and the exact RSET behaviour. 

To alleviate these discrepancies, we notice that the action~\eqref{Eq:AnomAction4D} has the structure $b_E \mathfrak E(\phi,\phi)+b_F \mathfrak F(\phi,\psi)$, where $\mathfrak E$ and $\mathfrak F$ both contain the kinetic bilinear $\mathcal K(\cdot,\cdot)$, whose variation produces the kinetic operator $\Delta_4$. The only difference is that $\mathfrak E$ contains a diagonal term, while $\mathfrak F$ contains an off-diagonal one. The form of $\mathfrak E$ and $\mathfrak F$ naturally leads us to consider a new term
\begin{equation}
   \mathfrak H(\psi,\psi)\equiv\frac{1}{2}\int\dd[4]x \sqrt{-g} \pqty{-\psi\Delta_4 \psi+F\psi}.
\end{equation}
Since $\mathfrak H$ is Weyl and diffeomorphism-invariant, the corresponding contribution to the ASET is traceless and covariantly conserved:
\begin{equation}
\label{Eq:Htensor}
    H^{\mu\nu}=\frac{2}{\sqrt{-g}}\fdv{\mathfrak H}{g_{\mu\nu}},\qquad H^{\mu}_{~\mu}=0,\quad \nabla_\mu H^{\mu\nu}=0.
\end{equation}
Because $\mathfrak H$ is Weyl invariant, it can be added to the action with an arbitrary coefficient $\gamma$ without spoiling the consistency of the method. Moreover, due to its form, the new term does not change the field equation for $\psi$.

However, including the new tensor $H_{\mu\nu}$ does change the trace-free part of the ASET, leading to different predictions. The correct coefficient of $\mathfrak H$ in the effective action can only be determined by a calculation in the complete theory. For our purposes, we will use $\gamma$ as a free parameter that can be adjusted to enhance the method. Since the addition of Weyl invariant pieces is arbitrary anyway, we choose to include $\mathfrak H$ as the simplest and most natural modification that allows some freedom in the AIEA. The validity of this choice will only be determined \textit{a posteriori}, once we make some predictions.
Therefore, the ASET we will work with is given by
\begin{equation}\label{Eq:ASET}
    T^{\mu\nu}_{\text{anom}}=b_E E^{\mu\nu}+b_F F^{\mu\nu}+\gamma H^{\mu\nu},
\end{equation}
where explicit expressions for these quantities can be found in Appendix~\ref{App:EFHTensors}.

It is worth noting that there is a value of the coefficient $\gamma$ which diagonalizes the kinetic term, thus reducing the number of propagating fields to a single one~\cite{Mottola2016}. In fact, choosing $\gamma=\frac{b_F^2}{b_E}\equiv \gamma_s$ (`s' stands for `single field'), we get
\begin{equation}
    b_E \mathcal K(\phi,\phi)+2b_F \mathcal K(\phi,\psi)+\frac{b_F^2}{b_E}\mathcal K(\psi,\psi)=b_E\mathcal K\pqty{\phi+\frac{b_F}{b_E}\psi,\phi+\frac{b_F}{b_E}\psi},
\end{equation}
which means that the only propagating combination is $\varpi\equiv \phi+\frac{b_F}{b_E}\psi$, while the other one does not propagate. Likewise, the source terms combine to give
\begin{equation}
    \pqty{E-\frac{2}{3}\square R}(b_E \phi+b_F\psi)+F\pqty{b_F\phi+\frac{b_F^2}{b_E}\psi}=\pqty{b_E\pqty{E-\frac{2}{3}\square R}+b_F F}\varpi.
\end{equation}
This proves that, despite the presence of two independent terms in the Weyl anomaly, it is possible to fully reproduce it using a single field.
If we extract $\gamma_s$ from $\gamma$, we can rewrite the action as
\begin{equation}
\label{Eq:AnomActionSingleField}
    S_{\text{anom}}[g;\varpi,\psi]
    =
    \frac{b_E}{2}\int d^{4}x\sqrt{-g}\left\{-\varpi\Delta_{4}\varpi+\frac{\mathcal A_4}{b_E}\varpi\right\}+\frac{\gamma-\gamma_s}{2}\int d^{4}x\sqrt{-g}\left\{-\psi\Delta_{4}\psi+F\psi\right\},
\end{equation}
showing that the addition of $\mathfrak H(\psi,\psi)$ with a generic $\gamma$ coefficient is equivalent to the $l_1$-dependent modification of the Riegert action suggested in \cite{BalbinotFabbriShapiro1999}, after an appropriate rescaling of the auxiliary fields and with the identification $\gamma-\gamma_s=-\gamma_s l_1$. The original Riegert action ($l_1=1$) corresponds to $\gamma=0$.

One could go even further and consider $\mathfrak H(\zeta,\zeta)$ for a new field $\zeta$, as previously proposed in \cite{AndersonMottolaVaulin2007}. This would certainly make the method more flexible in exchange for introducing more integration constants, thus making it even more ambiguous.
On the other hand, we checked that this idea does not resolve some of the shortcomings highlighted in~\cite{BalbinotFabbriShapiro1999} in non-conformally flat spacetimes. We will reserve further investigation for a later publication to determine whether the inclusion of a third field would allow the method to reproduce the features of the Unruh state.

As we already saw in the $1+1$ case, the ASET allows us to capture (some of) the defining features of the various vacuum states by specifying the asymptotic behaviour or demanding regularity of the corresponding ASET at horizons. This amounts to selecting different homogeneous solutions to their field equations, which is possible given the huge kernel of $\Delta_4$.
Adding an homogeneous solution to the on-shell value of the $T_{anom}^{\mu\nu}$ produces a new ASET that differs from the first one by a conserved, traceless tensor: it is in this contribution that one can see the ``state-dependence", i.e. the genuinely quantum non-local effects.

Several complications arise when we try to apply the AIEA method in $3+1$ dimensions. Unlike the two-dimensional case, four-dimensional black hole spacetimes are not conformally flat; luckily, there is a set of questions that we expect to be able to answer to a satisfying degree of reliability due to the universal near-horizon conformal invariance of black hole spacetime~\cite{Carlip2002,FabbriNavarro-Salas2005}.
Additionally, the presence of angular directions makes it so that covariant conservation and the trace relation~\eqref{Eq:AnomalyD} do not uniquely fix all the components of the RSET, not even in the simplest case considered here. Therefore, one needs to carefully consider in which cases the anomalous effective action approximation is reliable~\cite{Bardeen:2018gca}. Another, more practical complication stems from the fact that the field equations for the auxiliary fields become fourth-order~\eqref{Eq:EoMs4D}, thus increasing the number of integration constants required to fix a solution and making it harder to find global, analytic solutions. Unfortunately, the RN spacetime is not one of the rare exceptions in which we have analytic solutions, which forces us to obtain the ASET through a combination of analytic and numerical techniques. We will deal with this problem in Subsection~\ref{Subsec:Aux} but, before doing so, let us take a moment to carefully examine the relation between this method and Page's approximation.

\subsection{Comparison with Page's approximation}
\label{Subsec:Page}

A well-known method for obtaining the approximate stress-energy tensor of Weyl-invariant matter fields is Page-Brown-Ottwill (PBO) approximation, which was devised by Page~\cite{Page1982} and later extended by Brown and Ottewill~\cite{BrownOttewill1985, BrownOttewillPage1986}.

Let us briefly describe the idea of this method in a way that makes it easier to compare with the AIEA approximation.
As we already saw, if the matter sector is Weyl-invariant, the behaviour of the effective action under scaling is completely dictated by the anomaly. For a generic renormalization scheme, one finds\footnote{We slightly changed the notation of~\cite{BrownOttewillPage1986} to match with ours: $a\to b_F$, $b\to b_E$, $c\to b'-2/3(b_E+b_F)$}
\begin{equation}
\label{Eq:PageAnomaly}
    \Gamma[g]-\Gamma[e^{-2\omega}g]=b_F A[g;\omega]+b_E B[g;\omega]+\left(b'-\tfrac{2}{3}(b_E+b_F)\right)C[g;\omega]
\end{equation}
where $A$, $B$ and $C$ are known non-linear functionals. Their explicit expression is
\begin{align}
    A[g;\omega]&=\int\dd[4]x\sqrt{-g}\pqty{\omega F+\frac{2}{3}\,\mathcal R(\omega)},\\
    B[g;\omega]&=\int\dd[4]x\sqrt{-g}\pqty{\omega \pqty{E-\frac{2}{3}\square R}-2\omega\Delta_4\omega+\frac{2}{3}\,\mathcal R(\omega)},\\
    C[g;\omega]&=\int\dd[4]x\sqrt{-g}\,\mathcal R(\omega),
\end{align}
where $E$ and $F$ are given in~\eqref{Eq:EFdefinitions} and we defined
\begin{equation}
    \mathcal R(\omega)\equiv -\frac{1}{12}(R^2[g]-e^{-4\omega}R^2[e^{-2\omega}g])=(R+3(\square\omega-\abs{\nabla\omega}^2))(\square\omega-\abs{\nabla\omega}^2).
\end{equation}
Assuming that one can find a $\omega(g)$ such that the trace anomaly $\langle\hat T^\mu_{~\mu}\rangle[e^{-2\omega(g)}g]$ vanishes in the Weyl transformed spacetime $e^{-2\omega}g$, then one can derive the relation
\begin{equation}
    \langle\hat T^{\mu\nu}\rangle[e^{-2\omega(g)}g]=\langle\hat T^{\mu\nu}\rangle[g]-T_{PBO}^{\mu\nu}[g;\omega(g)],
\end{equation}
where $T^{\mu\nu}_{PBO}$ is obtained by varying the r.h.s. of~\eqref{Eq:PageAnomaly} with respect to the metric at fixed $\omega$, and then setting $\omega=\omega(g)$.
In the Weyl transformed spacetime, there are states with vanishing stress-energy tensor, and Page has argued that there should also be states for which the RSET is negligible~\cite{BrownOttewill1983}. If this is the case, one can set (either exactly, for the first class of states, or approximately, for the second one)
\begin{equation}
    \langle\hat T^{\mu\nu}\rangle[g]\approx T^{\mu\nu}_{PBO}[g;\omega(g)].
\end{equation}
Albeit nice and accurate, this derivation hides the true difficulty of the method, which is finding a solution to $\langle\hat T^\mu_{~\mu}\rangle[e^{-2\omega}g]=0$.

The AIEA method we adopt is based on the choice $b'=0$, which has the advantage that $\mathcal R(\omega)$ drops from the WZ action~\eqref{Eq:PageAnomaly}: the resulting action is quadratic in $\omega$, which is the reason why we were able to find linear equations for the auxiliary fields in Section~\ref{Subsec:3+1}.
Given the form of $\mathcal R(\omega)$, one can easily incorporate it in the AIEA method by adding $-\frac{b'}{12} R^2$ to the quantum effective action $\Gamma[g]$ --- correspondingly, one would reinstate the $b'\square R$ term in the trace anomaly~\eqref{Eq:TraceAnomalyin4D}.

However, the main difference between the AIEA and the PBO approximation is that, in the former, the fields are required to satisfy the field equations, but they are not directly connected to the vanishing of the trace anomaly in a conformally related spacetime --- this might be true for $\varpi$ in~\eqref{Eq:AnomActionSingleField}, but $\psi$ has no such meaning. In particular, if we were to augment the trace anomaly by a $b'\square R$ term, it would be clear that the anomaly cannot be cancelled by solving only linear field equations because
\begin{equation}
    \sqrt{-g^\sigma}\langle \hat T^\mu_{~\mu}\rangle[g^\sigma]=\sqrt{-g}\bqty{b_F F+b_E\pqty{E-\frac{2}{3}\square R+4\Delta_4\sigma}+b'\pqty{\square R-6D_4\sigma}},
\end{equation}
where $D_4$ is a non-linear differential operator that contains quadratic and cubic terms.\footnote{Explicitly $
D_4\sigma=\Delta_4\sigma-2(\square\sigma)^2+2\nabla_\mu\nabla_\nu\sigma(\nabla^\mu\nabla^\nu\sigma-2\nabla^\mu\sigma\nabla^\nu\sigma-G^{\mu\nu})+2R^{\mu\nu}\nabla_\mu\sigma\nabla_\nu\sigma-2\square\sigma\abs{\nabla\sigma}^2$.}
In Brown and Ottewill's paper~\cite{BrownOttewill1985}, where $b'=2/3(b_E+b_F)$, one looks for $\omega$ such that the $b_E$ and $b_F$ dependent terms cancel independently in the Weyl transformed metric $e^{-2\omega}g$, arriving at the equations\footnote{Given the non-linearity of the equations, it is not even clear whether this is possible for a generic spacetime.}
\begin{gather}
    F-\pqty{E-\frac{2}{3}\square R}-4\Delta_4\omega=0,\\
    F+\frac{2}{3}\square R+4D_4(-\omega)=0.
\end{gather}
While the first equation bears some resemblance to the auxiliary field equations (one could say that $\omega\sim \psi-\phi$), the second is highly non-linear, hence making PBO approximation different from --- or just very difficult to fit in --- the AIEA method when $b'\ne 0$.

\subsection{Auxiliary field solutions}
\label{Subsec:Aux}
As a first step towards a complete analysis, we
restrict in this work to auxiliary field configurations with spherical symmetry and whose radial and time dependence can be separated. For now, let us assume a radial dependence only. We will later introduce the time dependence to retrieve the Unruh state (see Section~\ref{Subsec:UnruhStates}).
The field equation for the radial-dependent part 
of the auxiliary field equations~\eqref{Eq:Conformalon4Dequations} reads
\begin{align}
\label{Eq:ConformalonRadialEq4D}
    &\phi''''+\frac{4 \left(r^2-r_- r_+\right)}{r (r-r_-) (r-r_+)}\phi'''+\frac{4 r^3 (r_-+r_+)-2 r^2 \left(r_-^2+4 r_- r_++r_+^2\right)+4 r_-^2 r_+^2}{r^2 (r-r_-)^2 (r-r_+)^2}\phi''+\nonumber\\&-\frac{4 (r (r_-+r_+)-2 r_- r_+)}{r^3 (r-r_-) (r-r_+)}\phi'=\frac{6 r^2 (r_-+r_+)^2-24 r r_- r_+ (r_-+r_+)+20 r_-^2 r_+^2}{r^4 (r-r_-)^2 (r-r_+)^2},\nonumber\\
    &\psi''''+\frac{4 \left(r^2-r_- r_+\right)}{r (r-r_-) (r-r_+)}\psi'''+\frac{4 r^3 (r_-+r_+)-2 r^2 \left(r_-^2+4 r_- r_++r_+^2\right)+4 r_-^2 r_+^2}{r^2 (r-r_-)^2 (r-r_+)^2}\psi''+\nonumber\\&-\frac{4 (r (r_-+r_+)-2 r_- r_+)}{r^3 (r-r_-) (r-r_+)}\psi'=\frac{6 (r (r_-+r_+)-2 r_- r_+)^2}{r^4 (r-r_-)^2 (r-r_+)^2}.
\end{align}
These equations, despite linear, are particularly hard to solve. Nevertheless, Fuchsian theory~\cite{Ince1956} allows us to compute the asymptotic expansion of the solutions centered around the (regular) singular points $\{r_-,r_+,\infty\}$.\footnote{The origin $r=0$ is another regular singular point but we are not interested in it.} To distinguish the coefficients corresponding to each expansion, we will use the superscripts $(+)$ for outer horizon, $(-)$ for inner horizon, and $(\infty)$ for radial infinity. For further clarity, we introduce the following associated parameters:
\begin{equation}
    (+):\;\epsilon_+\equiv \frac{r}{r_+}-1\,\qquad (-):\;\epsilon_-\equiv \frac{r}{r_-}-1\,\qquad (\infty):\; \epsilon_\infty\equiv \frac{r_+}{r}.
\end{equation}

Since we are interested in the study of semiclassical effects close to the inner/Cauchy horizon for states that are regular at the outer/event horizon, we will first look for a solution that is regular around the latter. By general results in Fuchsian theory of differential equations, we know that the fields admit the following asymptotic expansion close to $r_+$
\begin{subequations}
\label{Eq:HHstates}
\begin{align}
    \phi&=\sum_{n=0}^\infty \epsilon_+^n \pqty{\bar\phi^{(+)}_n+\tilde\phi^{(+)}_n\ln\abs{\epsilon_+}},\\
    \psi&=\sum_{n=0}^\infty \epsilon_+^n \pqty{\bar\psi^{(+)}_n+\tilde\psi^{(+)}_n\ln\abs{\epsilon_+}}.
\end{align}
\end{subequations}
Each of these series is uniquely determined, at least in a suitable neighbourhood, once four integration constant are given. For simplicity, we choose these two sets of free parameters to be $C_\phi^{(+)}\equiv\{\bar\phi^{(+)}_0, \bar\phi^{(+)}_1, \tilde\phi^{(+)}_0, \tilde\phi^{(+)}_1\}$ and $C_\psi^{(+)}\equiv\{\bar\psi^{(+)}_0, \bar\psi^{(+)}_1, \tilde\psi^{(+)}_0, \tilde\psi^{(+)}_1\}$. Expressions for the $n>1$ coefficients can be obtained by inserting the expansions~\eqref{Eq:HHstates} in Eqs.~\eqref{Eq:ConformalonRadialEq4D} and solving order by order in $\epsilon_+$.

While these expansions start to fail for radii which are too far from the outer horizon, it remains true that the actual solutions are uniquely fixed by four integration constants. 
Our strategy to determine the full field configurations is then the following. Starting from the expansions~\eqref{Eq:HHstates} at the event horizon, we (i) fix some of the coefficients $C_\phi^{(+)}$ and $C_\psi^{(+)}$ to enforce the ASET to be regular there (details below), (ii) numerically integrate the field equations~\eqref{Eq:ConformalonRadialEq4D} to cover every region (taking only a few sample values of the unfixed coefficients is sufficient, as we will see) and (iii) match the numerical solution with the corresponding asymptotic expansions at the Cauchy horizon and infinity.

At the Cauchy horizon, the field expansions read
\begin{subequations}
\label{Eq:ExpansionsInner}
\begin{align}
    \phi&=\sum_{n=0}^\infty \epsilon_-^n \pqty{\bar\phi_n^{(-)}+\tilde\phi_n^{(-)}\ln\abs{\epsilon_-}},\\
    \psi&=\sum_{n=0}^\infty \epsilon_-^n \pqty{\bar\psi_n^{(-)}+\tilde\psi_n^{(-)}\ln\abs{\epsilon_-}},
\end{align}
\end{subequations}
where, according to the outlined strategy, the coefficients $C_\phi^{(-)}\equiv\{\bar\phi^{(-)}_0, \bar\phi^{(-)}_1, \tilde\phi^{(-)}_0, \tilde\phi^{(-)}_1\}$ and $C_\psi^{(-)}\equiv\{\bar\psi^{(-)}_0, \bar\psi^{(-)}_1, \tilde\psi^{(-)}_0, \tilde\psi^{(-)}_1\}$ will be (numerically) determined in terms of $C_\phi^{(+)}$ and $C_\psi^{(+)}$. Likewise, at infinity:
\begin{subequations}
\label{Eq:ExpansionsInf}
\begin{align}
    \phi&=\sum_{n=0}^\infty \epsilon_\infty^n \pqty{\epsilon_\infty^{-2}\,\bar\phi^{(\infty)}_{n-2}+\tilde\phi^{(\infty)}_n\ln\epsilon_\infty},\\
    \psi&=\sum_{n=0}^\infty \epsilon_\infty^n \pqty{\epsilon_\infty^{-2}\,\bar\psi^{(\infty)}_{n-2}+\tilde\psi^{(\infty)}_n\ln\epsilon_\infty},
\end{align}
\end{subequations}
with coefficients $C_\phi^{(\infty)}\equiv\{\bar\phi^{(\infty)}_{-2}, \bar\phi^{(\infty)}_{-1}, \bar\phi^{(\infty)}_0, \bar\phi^{(\infty)}_1\}$ and $C_\psi^{(\infty)}\equiv\{\bar\psi^{(\infty)}_{-2}, \bar\psi^{(\infty)}_{-1}, \bar\psi^{(\infty)}_0, \bar\psi^{(\infty)}_1\}$ that are respectively determined in terms of $C^{(+)}_\phi$ and $C^{(+)}_\psi$.

In both cases, due to the linearity of the field equations, we know that the relation to $C^{(+)}_\phi$ and $C^{(+)}_\psi$ will be linear up to a shift (i.e. affine). This implies that we could represent it using a transition matrix --- which is the same for $\phi$ and $\psi$ --- and a shift vector. These additional ingredients, which are all that we need to determine the ASET in the entire spacetime, depend only on background geometry and, for dimensional reasons, can only be functions of the ratio $Q/M$.

Since we are interested in vacuum contributions at Cauchy horizons, we will not try to reproduce the Boulware state in the following because of its singular behaviour at the outer/event horizon. 
While this state would be the one relevant for the study of vacuum effects in horizonless stellar objects~\cite{Numajiri:2024qgh}, in the next subsections we will consider instead the Hartle-Hawking and Unruh states.

\subsection{Hartle-Hawking state}
\label{Subsec:HHStates}

\subsubsection{Behaviour at the event and Cauchy horizons}
In order to ensure regularity at the event horizon for a static state, we need to make sure that $T_{uu}=T_{vv}\sim\order{r-r_{+}}^{2}$ and $T_{uv}\sim\order{r-r_{+}}^1$. One can check that this amounts to impose ten conditions,\footnote{Explicitly, in terms of $\epsilon_+\equiv\frac{r}{r_+}-1$, one needs to kill the $\epsilon_+ ^{-3}, \epsilon_+ ^{-2}, \epsilon_+ ^{-1}, \epsilon_+ ^{-1}\ln \epsilon_+ ,\ln \epsilon_+ , (\ln \epsilon_+ )^2$  divergences in $T_{uu}/f^2$, and the $\epsilon_+ ^{-2}, \epsilon_+ ^{-1}, \ln \epsilon_+ , (\ln \epsilon_+ )^2$ divergences in $T_{uv}/f$. \label{Ftn:HHHorizonRegularity}} which, given the form of the ASET, reduce to just five.
As we show in the Appendix~\ref{App:Regularity Conditions}, these conditions are satisfied if the fields themselves are regular there\footnote{There exist non-regular field solutions which combine to produce a regular ASET nonetheless. However, we discard them as they lead to incorrect predictions for the ASET values at the event horizon.}
\begin{equation}
\label{Eq:HHMiminamlSolution}
    \text{M:}\quad \tilde\phi^{(+)}_0=\tilde\phi^{(+)}_1=\tilde\psi^{(+)}_0=\tilde\psi^{(+)}_1=0.
\end{equation}
In this \textit{minimal} solution, the remaining coefficients (i.e. $\bar\phi^{(+)}_0$, $\bar\phi^{(+)}_1$, $\bar\psi^{(+)}_0$, $\bar\psi^{(+)}_1$) are completely free and can be adjusted to reproduce some of the expected feature of a Hartle-Hawking state. As we show below, we can fix the parameters $\bar\phi^{(+)}_1$ and  $\bar\psi^{(+)}_1$ to construct a state that has the correct singular behaviour of $T_{uu}=T_{vv}\sim \order{r-r_{-}}^{0}$ at the Cauchy horizon, and also describes, at leading order, a 4D thermal bath at the Hawking temperature at infinity.

The leading-order behaviour in the ASET at the Cauchy horizon is obtained by replacing the expansions~\eqref{Eq:ExpansionsInner} in the components of the ASET (included in Appendix~\ref{App:EFHTensors}) and expanding in series once again. Using $(t,r,\theta,\phi)$ coordinates for convenience, we identify the following divergent contributions at the Cauchy horizon
\begin{equation}
    \begin{split}
        \label{Eq:ASETInner}
T^{\mu}_{~\nu}
&
={A^{(-)}}^{\mu}_{~\nu}\,\epsilon_-^{-2}+{B^{(-)}}^{\mu}_{~\nu}\,\epsilon_-^{-1}
+{C^{(-)}}^{\mu}_{~\nu}\ln^2\abs{
\epsilon_-}
+{D^{(-)}}^{\mu}_{~\nu}\ln\abs{\epsilon_-}+\order{\epsilon_-}^0.
    \end{split}
\end{equation}
The coefficient of the leading-order divergence is
\begin{equation}
    {A^{(-)}}^{\mu}_{~\nu}=\frac{2\kappa_{-}^{2}}{r_{-}^{2}}\q{\tilde{\phi}_{0}^{(-)},\tilde{\psi}_{0}^{(-)}}\text{diag}\left(-1,\frac{1}{3},\frac{1}{3},\frac{1}{3}\right),
\end{equation}
where we have introduced the quadratic form
\begin{equation}\label{Eq:QuadForm}
\q{x,y}=b_{E}x^{2}+2b_{F}xy+\gamma y^2.
\end{equation}
The next contribution has the slightly more involved form
\begin{align}\label{Eq:BmunuTensor}
    {B^{(-)}}^{\mu}_{~\nu}
    &
    =\frac{2\kappa_{-}}{3r_{-}^{3}}\pqty{\frac{2r_+}{r_-}-1}\q{\tilde{\phi}_{0}^{(-)},\tilde{\psi}_{0}^{(-)}}\text{diag}\left(-3,5,-1,-1\right)\nonumber\\
    &
    +\frac{4\kappa_{-}}{3r_{-}^{4}}\pqty{b_E(\kappa_-r_-^2\tilde{\phi}_{1}^{(-)}-3r_+\tilde{\phi}_{0}^{(-)})+b_F(\kappa_-r_-^2\tilde{\psi}_{1}^{(-)}-3r_+\tilde{\psi}_{0}^{(-)})}\text{diag}\left(-1,1,0,0\right)\nonumber\\
    &
    -\frac{8\kappa_{-}^2}{3r_{-}^{2}}\dif{\tilde\phi_{0}^{(-)},\tilde\phi_{1}^{(-)}+\frac{3}{2},\tilde\psi_{0}^{(-)},\tilde\psi_{1}^{(-)}+\frac{3}{2}}\text{diag}\left(-1,1,0,0\right),
\end{align}
where we used the short-hand notation $
    \dif{x,y,z,w}\equiv \q{x+y,z+w}-\q{x,z}-\q{y,w}$.
This is followed by
\begin{equation}
    {C^{(-)}}^{\mu}_{~\nu}=\frac{4\kappa_{-}^{2}}{r_{-}^{2}}\q{\tilde{\phi}_{1}^{(-)},\tilde{\psi}_{1}^{(-)}}\text{diag}\left(1,1,-1,-1\right),
\end{equation}
which controls the leading (quadratic) logarithmic divergence, and 
\begin{equation}
\begin{split}
{D^{(-)}}^{\mu}_{~\nu}
&
=\frac{4\kappa_{-}^{2}}{3r_{-}^{2}}\q{\tilde{\phi}_{1}^{(-)},\tilde{\psi}_{1}^{(-)}}\,\text{diag}\left(-1,-5,3,3\right)+\\
&
-\frac{4\kappa_{-}r_+}{r_{-}^{4}}\pqty{b_F \tilde\phi_{1}^{(-)}+\gamma \tilde\psi_{1}^{(-)}}\text{diag}\left(-1,-1,1,1\right)+\\
&-\frac{2\kappa_-(3r_--2r_+)}{3r_-^4}\dif{\tilde\phi_{0}^{(-)}-\frac{3}{2},\tilde\phi_{1}^{(-)},\tilde\psi_{1}^{(-)}-\frac{3}{2},\tilde\psi_{1}^{(-)}}\text{diag}\left(-1,-1,1,1\right)+\\&+\frac{4\kappa_-^2}{3r_-^2}\dif{\bar\phi_{1}^{(-)}-9,\tilde\phi_{1}^{(-)},\bar\psi_{1}^{(-)}-9,\tilde\psi_{1}^{(-)}}\text{diag}\left(-1,-1,1,1\right),
\end{split}
\end{equation}
which controls the sub-leading (linear) logarithmic one.

It is easy to see that all these tensors are traceless, ensuring the finiteness of the trace anomaly at the Cauchy horizon.
From the above expressions, we can easily examine the regularity conditions in double-null Eddington-Finkelstein coordinates~\eqref{Eq:Regularity-}:
\begin{align}\label{Eq:DivTuuHH}
    T_{uu}
    &=T_{vv}=\frac{f}{4}\left(T^{r}_{~r}-T^{t}_{~t}\right)
    =\frac{\kappa_{-}r_{-}}{2}\left[\left(A^{r}_{~r}-A^{t}_{~t}\right)\epsilon_-^{-1}+B^{r}_{~r}-B^{t}_{~t}+\order{\epsilon_-}\right],\\
    T_{uv}
    &=\frac{f}{4}\left(T^{r}_{~r}+T^{t}_{~t}\right)
    =\frac{\kappa_{-}r_{-}}{2}\left[\left(A^{r}_{~r}+A^{t}_{~t}\right)\epsilon_-^{-1}+B^{r}_{~r}+B^{t}_{~t}+\order{\epsilon_-}\right].
\end{align}

Recall that the freedom to choose the field boundary conditions allows, in principle, the construction of ASETs with asymptotic behaviours that do not correspond to any of the well-known states. Equation~\eqref{Eq:DivTuuHH} shows that the ASET fluxes generically diverge at the Cauchy horizon, a behaviour which has not been observed in exact numerical calculations for minimally and conformally coupled scalar fields, neither for the Hartle-Hawking~\cite{Sela:2018xko,Zilberman:2019buh} nor the Unruh states~\cite{Sela:2018xko,Hollands:2019whz,Hollands:2020qpe,Zilberman:2019buh,Klein:2023urp}. Therefore, to better approximate the true Hartle-Hawking state, we need to eliminate this spurious divergence by imposing the condition 
\begin{equation}\label{Eq:InnerHorCond}
\q{\tilde\phi_{0}^{(-)},\tilde\psi_{0}^{(-)}}=0.
\end{equation}
This is achieved by fine tuning the free coefficients $\bar\phi_{1}^{(+)}$ and $\bar\psi_{1}^{(+)}$ so that the numerically determined $\bar\phi_{1}^{(-)}$ and $\bar\psi_{1}^{(-)}$ satisfy~\eqref{Eq:InnerHorCond}. 
By inspection of the coefficient of the next order in the expansion, i.e., Eq.~\eqref{Eq:BmunuTensor}, we see that imposing condition~\eqref{Eq:InnerHorCond} implies that the energy density and radial pressure in the ASET diverge as $\propto \epsilon_{-}^{-1}$, while the angular pressures have a weaker divergence that is $\propto\ln^{2}\abs{\epsilon_{-}}$ at most.

The ASET expansion~\eqref{Eq:ASETInner} reveals the hierarchy of divergent terms that an RSET might contain. Cancelling all these divergences would require satisfying four independent conditions, which are $A^{\mu}_{~\nu}{}^{(-)}=B^{\mu}_{~\nu}{}^{(-)}=C^{\mu}_{~\nu}{}^{(-)}=D^{\mu}_{~\nu}{}^{(-)}=0$. We have found that no combination of the remaining free coefficients in the minimal solution~\eqref{Eq:HHMiminamlSolution} can satisfy all four simultaneously. Therefore, among the broad space of ``states" that can be described with the AIEA method, we find none that is regular at the event and Cauchy horizons. This result is consistent with standard analyses of vacuum states in black hole spacetimes. 

Exact calculations are so technically challenging that subleading (logarithmic) divergences in the RSET have been conjectured to exist, but have not been calculated explicitly. Hence, we cannot state with certainty whether the logarithmic terms in~\eqref{Eq:ASETInner} are genuine features or artifacts of the approximation we use. Since we only aim to reproduce leading-order divergences, we impose no additional constrain on the ASET at the Cauchy horizon. Nonetheless, to guarantee that the ASET provides the best possible approximation to the RSET in the true Hartle-Hawking state, we need to examine its properties at infinity.

\subsubsection{Thermality at infinity: Page's approximation}
The Hartle-Hawking state describes a thermal bath of particles at the Hawking temperature, so its RSET must approach the following form at large distances
\begin{equation}
\label{Eq:ThermalRadiation4D}
    \langle\hat T^\mu{}_\nu\rangle\underset{r\to\infty}{\sim}\rho(T)\,\text{diag}\pqty{-1,\frac{1}{3},\frac{1}{3},\frac{1}{3}},
\end{equation}
where $\rho(T)$ is the radiation energy density, which scales as $T^4$ in 4D with a coefficient that depends on the number of fields~\cite{CandelasHoward1984, Grovesetal2002, Jensen:1988rh}
\begin{equation}
\label{Eq:FVThermalBath}
    \rho(T)=\pqty{N_s+\frac{7}{2} N_f+N_v}\frac{\pi^2 T^4}{30}.
\end{equation}

Let us now look at the form of the ASET for $r\to\infty$, where $T$ approaches the Hawking temperature $\kappa_+/2\pi$. Replacing the expansions~\eqref{Eq:ExpansionsInf} in the ASET components, expressed in $(t,r,\theta,\phi)$ coordinates for convenience, we obtain the leading-order behaviour
    \begin{equation}\label{Eq:ASETInfHH}
    T^{\mu}_{~\nu}=\frac{14}{r_{+}^{4}}\q{\bar{\phi}^{(\infty)}_{-2},\bar{\psi}^{(\infty)}_{-2}}~\text{diag}\left({-1,\frac{1}{3},\frac{1}{3},\frac{1}{3}}\right)+\order{\frac{1}{r}},
    \end{equation}
This expression tells us that, if we want the ASET to be a good approximation of the full RSET at infinity, we must require $\gamma\ge \gamma_s$. If $\gamma$ were smaller, the quadratic form~\eqref{Eq:QuadForm} would be negative-definite and we could not have a positive energy density. This restricts the space of physically acceptable AIEA approximations to those with $\gamma\ge \gamma_s$.
Within this range of values, we exploit the freedom in adjusting the field integration constants to impose that 
\begin{equation}\label{Eq:InfinityCond}
\q{\bar{\phi}^{(\infty)}_{-2},\bar{\psi}^{(\infty)}_{-2}}=\frac{r_{+}^{4}}{14}\rho\pqty{\frac{\kappa_+}{2\pi}}.
\end{equation}

However, the Hartle-Hawking state is also an equilibrium state, which means that the local temperature must redshift with distance according to Tolman's law. This implies that, up to order $1/r^5$, all the coefficients of the RSET asymptotic expansion are fixed. Beyond this order, the trace anomaly kicks in, spoiling the perfect gas form~\eqref{Eq:ThermalRadiation4D}. In the case of Schwarzschild spacetime, it was noticed in~\cite{BalbinotFabbriShapiro1999} that the solution intended to represent the Hartle-Hawking state does not match with said expansion. Here, we show that this observation is more general, as the ASET fails to comply with Tolman's law even off-shell of the auxiliary fields. Indeed, one can rapidly check that, for any auxiliary field configuration, the coefficients $\varrho_n$ of $\epsilon_\infty^n$ in the expansion of the ASET $T^t_{~t}$ component at infinity satisfy
\begin{equation}
    \varrho_3-\frac{r_-+r_+}{r_+} \varrho_2-\frac{r_-^2-18r_-r_++r_+^2}{24\,r_+^2}\varrho_1-\frac{(r_-+r_+)(39r_-^2+2r_-r_++39r_+^2)}{252\,r_+^3}\varrho_0=0,
\end{equation}
which is a relation that just does not hold for the RSET~\eqref{Eq:ThermalRadiation4D}. 

It is interesting to compare the ASET with the stress-energy tensor obtained in other approximation schemes. Huang's evaluation of the RSET~\cite{Huang1992} shows that Page's approximation is able to reproduce the expected thermal behaviour~\eqref{Eq:ThermalRadiation4D} at large distances. However, at the Cauchy horizon, it returns the following components in $(t,r,\theta,\phi)$ coordinates
\begin{equation}
    T^{\mu}_{~\nu}=-\frac{\kappa_{-}^{3}\left(r_{+}+r_{-}\right)\left(r_{+}^{2}+r_{-}^{2}\right)\left(r_{+}^{4}+r_{-}^{4}\right)}{960\pi^{2}r_{+}^{8}}\epsilon_-^{-2}~\text{diag}\left(-1,\frac{1}{3},\frac{1}{3},\frac{1}{3}\right)+\order{\epsilon_-}^{-1}.
\end{equation}
Since these are incompatible with the constant $T_{uu}$ and $T_{vv}$ fluxes that were found numerically, Huang's approximation can be deemed inadequate at the Cauchy horizon. On the other hand, the Anderson-Hiscock-Samuel approximate RSET~\cite{Andersonetal1995} shows a logarithmic divergence at the event horizon of a non-extremal Reissner-Nordstr{\"o}m black hole --- which becomes a combination of linear and logarithmic divergences for extremal black holes --- signalling that this approximation cannot be extended towards the interior in any case.

As we prove below, the AIEA method is the only known approximation that yields a 4D RSET approximation for the Hartle-Hawking state with the correct (regular) behaviour at the event horizon, the right leading-order divergence at the Cauchy horizon, and which approaches a thermal bath at infinity. 

\subsubsection{Numerical results: scalar, fermion and vector fields}
As already mentioned, we leverage the numerical relations between the coefficients of the field expansion around different points to convert the conditions~\eqref{Eq:InnerHorCond} and~\eqref{Eq:InfinityCond} into (numerical) conditions for $\bar\phi_{1}^{(+)}$ and $\bar\psi_{1}^{(+)}$. We do so in what follows to generate numerical results for the ASET at the Cauchy horizon.
Given the initial conditions at the event horizon
\begin{equation}
    C^{(+)}_\phi=\left\{\bar\phi^{(+)}_0,\bar\phi^{(+)}_1,0,0\right\},\quad C^{(+)}_\psi=\left\{\bar\psi^{(+)}_0,\bar\psi^{(+)}_1,0,0\right\},
\end{equation}
we numerically integrate the field equations to extend the solutions everywhere, following the strategy outlined in Subsec.~\ref{Subsec:Aux}. In particular, we are able to numerically determine the following relations for the coefficients at the inner horizon
\begin{subequations}
\label{Eq:HHRelsInner}
\begin{align}
    \tilde\phi^{(-)}_0&=M_{\phi}^{(-)} \bar\phi^{(+)}_1+d^{(-)}_{\phi},\qquad \tilde\phi^{(-)}_1=M'{}^{(-)}_{\phi} \bar\phi^{(+)}_1+d'{}^{(-)}_{\phi},\\
    \tilde\psi^{(-)}_0&=M^{(-)}_{\psi} \bar\psi^{(+)}_1+d^{(-)}_{\psi},\qquad \tilde\psi^{(-)}_1=M'{}^{(-)}_{\psi} \bar\psi^{(+)}_1+d'{}^{(-)}_{\psi},
\end{align}
\end{subequations}
and for those at infinity
\begin{subequations}
\label{Eq:HHRelsInf}
    \begin{align}
    \bar\phi^{(\infty)}_{-2}&=M^{(\infty)}_{\phi} \bar\phi^{(+)}_1+d^{(\infty)}_{\phi},\\
    \bar\psi^{(\infty)}_{-2}&=M^{(\infty)}_{\psi} \bar\psi^{(+)}_1+d^{(\infty)}_{\psi}.
\end{align}
\end{subequations}
As expected, all the $M_{\phi/\psi}$, $M'_{\phi/\psi}$, $d_{\phi/\psi}$, and $d'_{\phi/\psi}$ are functions of $Q/M$ only. Moreover, since the homogeneous part of the field equation is the same for both fields, one has $M^{(-)}_\phi=M^{(-)}_\psi$, $M'{}^{(-)}_\phi=M'{}^{(-)}_\psi$, and $M^{(\infty)}_\phi=M^{(\infty)}_\psi$.
Notice that the free coefficients $\bar\phi^{(+)}_0$ and $\bar\psi^{(+)}_0$ do not contribute to the relations~\eqref{Eq:HHRelsInner} and~\eqref{Eq:HHRelsInf} so they do not affect the leading-order of the ASET at the Cauchy horizon and at infinity. On the other hand, they modify the sub-leading behaviour in the region between the horizons and in the bulk of the exterior region, and can thus be adjusted to make the ASET match the exact RSET more accurately. Here, since we are only interested in the asymptotic values, we fix $\bar\phi^{(+)}_0=0$ and $\bar\psi^{(+)}_0=0$ without loss of generality.
 
Since the above matrices and vectors only depend on $Q/M$, it is enough to calculate them once for a wide range of values of $Q/M$ to be able to generate ASET results for any field content and $\gamma$ value. We solve the field equations in the range $r_{-}<r<r_{\infty}$, where $r_{\infty}\gg r_{+}$ is a sufficiently large fiducial radius where the ASET is well-approximated by the expansion~\eqref{Eq:ASETInfHH}. For generic values of $\gamma>\gamma_{s}$, we find two sets of $\left\{\bar\phi^{(+)}_1,\bar\psi^{(+)}_1\right\}$ values that satisfy conditions~\eqref{Eq:InnerHorCond} and~\eqref{Eq:InfinityCond} simultaneously. 
We explored a wide range of $\gamma$ values and found that the ASET shows good agreement with exact numerical results for minimal fields~\cite{Zilberman:2022aum} as $\gamma\to\gamma_{s}$. However, we cannot directly take $\gamma=\gamma_s$ because the two families of $\left\{\bar\phi^{(+)}_1,\bar\psi^{(+)}_1\right\}$ values would degenerate (recall that there is a single propagating field for $\gamma=\gamma_s$). In this sense, the procedure has a discontinuous limit.
For clarity, we present here the results for $\gamma-\gamma_s=10^{-6}$ only, a case in which the two sets of $\left\{\bar\phi^{(+)}_1,\bar\psi^{(+)}_1\right\}$ values are almost indistinguishable, and plot just one of these sets.

\begin{figure}
    \centering
    \includegraphics[width=0.7\linewidth]{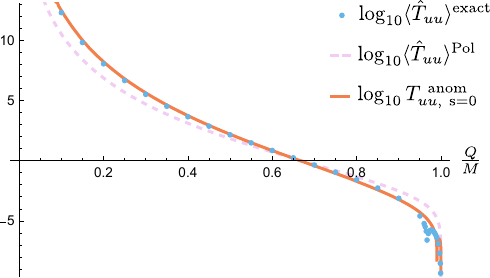}
    \caption{Flux components of the ASET at the Cauchy horizon (single scalar field) in terms of $Q/M$. In orange we show the values of $\log_{10}{\left|T_{uu}\right|}$ obtained through the AIEA prescription, which are positive until $Q/M\approx0.99$, and vanish in the extremal limit. The ASET shows a better agreement with the exact values obtained for minimally coupled fields~\cite{Zilberman:2019buh} (blue dots) than the Polyakov approximation (dashed pink curve), which predicts $T_{uu}=(\kappa_{+}^{2}-\kappa_{-}^{2})/192\pi^{2}r_{-}^2<0$ for all $Q/M$ values instead.}
    \label{Fig:HHRSETLogScalar}
\end{figure}

Figure~\ref{Fig:HHRSETLogScalar} shows the ASET component $\log_{10}\left|T_{uu}\right|=\log_{10}\left|T_{vv}\right|$ for a single scalar field evaluated exactly at $r=r_{-}$ as a function of $Q/M$. Our results are contrasted with the exact numerical data obtained for minimally coupled field in~\cite{Zilberman:2019buh}, and with the Polyakov approximation~\eqref{Eq:Polyakov}.\footnote{Results for the Polyakov approximation at the Cauchy horizon are obtained by multiplying the flux components~\eqref{Eq:HHFluxes2D} by the factor $1/4\pi r_{-}^{2}$.} We are comparing with minimal fields because the data is publicly available~\cite{Zilberman:2019buh}, but results for the $\langle \hat{T}_{vv}\rangle$ component of conformally coupled fields in the Unruh state can be consulted in~\cite{Hollands:2019whz,Klein:2023rwg} (recall that the Unruh and Hartle-Hawking fluxes just differ by the Hawking outflux term, which is small and constant). 
From a review of the literature, we infer that the exact Hartle-Hawking RSET for conformal fields will exhibit constant positive fluxes at the Cauchy horizon for small and intermediate $Q/M$ values, and transition to negative fluxes for $Q/M$ values near extremality. These characteristics are well-reproduced by the ASET. 

\begin{figure}
    \centering
    \includegraphics[width=0.7\linewidth]{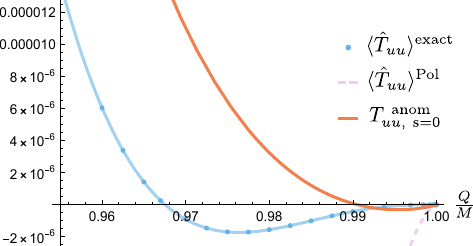}
    \caption{Flux components $T_{uu}$ of the ASET at the Cauchy horizon (single scalar field) for values of $Q/M$ close to extremality. The values computed through the AIEA method (orange) change sign at $Q/M\approx0.99$ and vanish in the extremal limit. However, in this range of $Q/M$ values they are not as good of an approximation to the values obtained numerically (blue dots). The Polyakov approximation, also negative near extremality, is shown in dashed pink. }
    \label{Fig:HHRSETScalar}
\end{figure}

For small $Q/M$ values, the ASET diverges in the $Q/M\to0$ limit towards $+\infty$, in contrast to the divergence towards $-\infty$ seen in Polyakov's approximation.
For intermediate $Q/M$ values, the AIEA method shows excellent agreement, both in sign and magnitude, with the exact results. The difference in slope between the Polyakov and AIEA approximations is due to the 4D fluxes not being proportional to the difference between the surface gravities squared, as in~\eqref{Eq:HHFluxes2D}. For values of $Q/M$ near extremality, shown in detail in Fig.~\ref{Fig:HHRSETScalar}, the approximation become less accurate, but still exhibits a change in sign near $Q/M\approx0.99$. This sign change is a feature of the RSET in the near-extremal limit of Reissner-Nordstr{\"o}m~\cite{Zilberman:2019buh} and Kerr~\cite{Zilberman:2022aum} black holes which the Polyakov approximation fails to reproduce.

\begin{figure}
    \centering
    \includegraphics[width=0.7\linewidth]{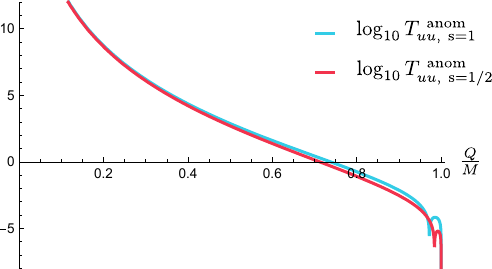}
    \caption{Flux components of the ASET at the Cauchy horizon in terms of $Q/M$ for single fermion (red) and vector (blue) fields. We are showing the values of $\log_{10}{\left|T_{uu}\right|}$ obtained through the AIEA prescription, which are positive until $Q/M\approx0.972$ for the vector field and until $Q/M\approx0.984$ for the fermion one. The ASET at the Cauchy horizon is very close in magnitude to the results for the scalar field shown in Fig~\ref{Fig:HHRSETLogScalar}, hence we conjecture they should resemble the ones obtained via exact calculations for conformal fields as well.}
    \label{Fig:HHFermionVectorLog}
\end{figure}

\begin{figure}
    \centering
    \includegraphics[width=0.7\linewidth]{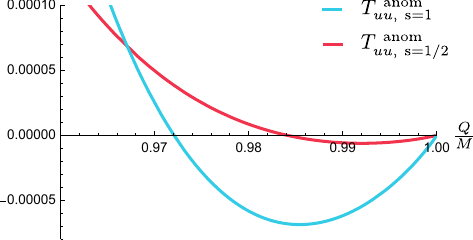}
    \caption{Flux components $T_{uu}$ of the ASET at the Cauchy horizon (single scalar field) for values of $Q/M$ close to extremality. These fluxes are positive for $Q/M<0.972$ for the vector field and for $Q/M<0.984$ for the fermion one. }
    \label{Fig:HHFermionVector}
\end{figure}

Results for the single scalar field approximate well the numerical results from~\cite{Zilberman:2019buh}. As the AIEA method is also valid for conformal fermion and vector fields, we have generated numerical results for them as well. Figure~\ref{Fig:HHFermionVectorLog} shows the logarithm of the flux component $T_{uu}$ for a single fermion and vector fields. To obtain them, we matched the ASET with the corresponding values~\eqref{Eq:FVThermalBath} for a thermal bath asymptotically. We obtain qualitatively similar results, with positive fluxes that decrease as $Q/M$ increases, eventually changing sign near extremality, as depicted in Fig.~\ref{Fig:HHFermionVector}.

We have shown that the AIEA prescription provides can be used to obtain an ASET that reproduces the main features of the Hartle-Hawking state. This ASET reproduces the main asymptotic properties of the exact RSET: its regular at the event horizon, has finite energy fluxes at the Cauchy horizon and describes a thermal bath at the Hawking temperature at infinity. This is a great success for the AIEA method. In the next Subsection we analyze the more complicated case of attempting to describe Unruh-like states, where we will be faced with the limitations in the AIEA prescription.  

\subsection{Unruh state}
\label{Subsec:UnruhStates}
Learning from the lesson of the $1+1$ case, we see that a flux term $T^r_{~t}$ can be easily incorporated in the ASET without making it time-dependent, just by adding a linear time dependence in the auxiliary fields~\eqref{Eq:1+1ASET}.
In a static spherically symmetric spacetime, the radial dependence of such a component is fixed by covariant conservation:
\begin{equation}
    \nabla_\mu T^\mu_{~t}=0 \implies T^r_{~t}=-\frac{L}{4\pi r^2},
\end{equation}
with $L$ being the so-called (integrated) luminosity.

In $3+1$ we have to deal with the slight complication that, in a non-Ricci, non-Weyl flat spacetime (such as Reissner-Nordstr\"om), the ASET contains an explicit dependence on the field values through terms of the form
\begin{subequations}
\label{Eq:plret}
\begin{align}
    F^{\mu\nu}&\supset-4\nabla^\rho\nabla^\sigma C^{(\mu}{}_\rho{}^{\nu)}{}_\sigma\,\phi-2C^{\mu}{}_\rho{}^{\nu}{}_\sigma R^{\rho\sigma}\phi,\\
    H^{\mu\nu}&\supset-4\nabla^\rho\nabla^\sigma C^{(\mu}{}_\rho{}^{\nu)}{}_\sigma\,\psi-2C^{\mu}{}_\rho{}^{\nu}{}_\sigma R^{\rho\sigma}\psi.
\end{align}
\end{subequations}
As mentioned, the only allowed time dependence is in the linear shifts $\phi\to\phi(r)+p_\phi t$, $\psi\to\psi(r)+p_\psi t$. With no additional constraints, this would translate into a $t$-dependent ASET. However, given that the functional dependence on $\phi$ and $\psi$ is the same~\eqref{Eq:plret}, we can recover a $t$-independent ASET by requiring 
\begin{equation}
    \label{Eq:TimeIndependenceASET}
    b_F p_\phi+\gamma p_\psi=0.
\end{equation}
While we cannot exclude that other, more complicated forms of time dependence in the field could still result in a time-independent ASET, these solutions would be hard to find explicitly, and would also make the field equations into partial differential equations in two variables, spoiling the simplicity of the treatment.

Since the Unruh state is not consistent with the symmetries of the background spacetime, we cannot ask for regularity on both branches of the event horizon. This state is the one approached by the physical \textit{in} vacuum at late times in a gravitational collapse. Hence, we can safely disregard any issue related to divergences on the past branch and require regularity on the future branch only, yielding the regularity conditions~\eqref{Eq:Regularity+}. 
As showed in the Appendix~\ref{App:Regularity Conditions}, we can achieve regularity at the future horizon with the \textit{minimal} solution
\begin{equation}
\label{UnruhMinimalSolution}
    \text{M:}\quad \tilde\phi^{(+)}_0=\frac{p_\phi}{2\kappa_+},\quad\tilde\psi^{(+)}_0=\frac{p_\psi}{2\kappa_+},\quad\tilde\phi^{(+)}_1=\tilde\psi^{(+)}_1=0,
\end{equation}
which reduces to~\eqref{Eq:HHMiminamlSolution} as $p_\phi, p_\psi\to 0$. As mentioned, $p_\phi$ and $p_\psi$ cannot be chosen independently though, meaning that the resulting solution has only $5$ free parameters: namely $\bar\phi^{(+)}_0$, $\bar\phi^{(+)}_1$, $\bar\psi^{(+)}_0$, $\bar\psi^{(+)}_1$, and $p_\psi$.

Following a similar construction to Subsec.~\ref{Subsec:HHStates}, we use the asymptotic expansions~\eqref{Eq:HHstates} with the minimal solution~\eqref{UnruhMinimalSolution}, and integrate the fields toward infinity and the Cauchy horizon. In these regions, we need to impose certain conditions to the ASET to be certain that we are selecting the ``true" Unruh state. At the Cauchy horizon, for consistency with numerical results~\cite{Hollands:2019whz,Zilberman:2019buh,Klein:2023urp}, we need to enforce that the ASET has finite fluxes:
\begin{subequations}
\begin{align}
    T_{uu}
    &
    =\frac{f}{4}\left(T^{r}_{~r}-T^{t}_{~t}+\frac{2T^{r}_{~t}}{f}\right)=\order{r-r_{-}}^{0},\\
    T_{vv}
    &
    =\frac{f}{4}\left(T^{r}_{~r}-T^{t}_{~t}-\frac{2T^{r}_{~t}}{f}\right)=\order{r-r_{-}}^{0}.
\end{align}
\end{subequations}
These are achieved by imposing the condition
\begin{equation}
\label{Eq:UnruhInnerRegular}
    \gamma (\gamma_{s}-\gamma)p_{\psi}^{2}+4\gamma_s \kappa_-^2 \q{\tilde\phi^{(-)}_{0},\tilde\psi^{(-)}_{0}}=0,
\end{equation}
where we have used again the quadratic form defined in~\eqref{Eq:QuadForm}.
Since distant observer at $\mathscr I^+$ should perceive the Hawking fluxes~\cite{Candelas1980}, which requires (in $(t,r,\theta,\varphi)$ coordinates)
\begin{equation}
    T^{\mu}_{~\nu}\overset{r\to\infty}{\sim}\order{\frac{1}{r^{2}}},
\end{equation}
we also want to enforce
\begin{equation}
\label{Eq:UnruhInfinityRegular}
    \begin{cases}
        \q{
\bar\phi^{(\infty)}_{-2},\bar\psi^{(\infty)}_{-2}} = 0,\\
    b_E \bar\phi^{(\infty)}_{-1}\bar\phi^{(\infty)}_{-2}+b_F\bar\phi^{(\infty)}_{-1}\bar\psi^{(\infty)}_{-2}+b_F \bar\psi^{(\infty)}_{-1}\bar\phi^{(\infty)}_{-2}+\gamma \bar\psi^{(\infty)}_{-1}\bar\psi^{(\infty)}_{-2}=0,
    \end{cases}
\end{equation}
to eliminate constant (cfr. Eq.~\eqref{Eq:ASETInfHH}) and $\order{1/r}$ terms at infinity.
Additionally, the condition that the state is vacuum at $\mathscr I^-$ requires
\begin{equation}
    r^2 T_{vv}\overset{r\to\infty}{\to} 0
\end{equation}
which, in terms of the field coefficients, can be implemented by
\begin{equation}
    \begin{split}
    \label{Eq:UnruhInfinityVacuum}        &2\pqty{1+\frac{r_-}{r_+}}\dif{\tilde\phi^{(\infty)}_{-2}+\frac{p_\phi}{2\kappa_+},\frac{p_\phi}{2\kappa_+},\tilde\psi^{(\infty)}_{-2}+\frac{p_\psi}{2\kappa_+},\frac{p_\psi}{2\kappa_+}}+\q{\frac{p_\phi}{2\kappa_+},\frac{p_\psi}{2\kappa_+}}+\\&+\frac{r_-+r_+}{r_--r_+}\q{\tilde\phi^{(\infty)}_{-1},\tilde\psi^{(\infty)}_{-1}}-3\q{\tilde\phi^{(\infty)}_{-1}+\frac{p_\phi}{2\kappa_+},\tilde\psi^{(\infty)}_{-1}+\frac{p_\psi}{2\kappa_+}}=0.
    \end{split}
\end{equation}

One could proceed by calculating the matrix and vector components, analogous to~\eqref{Eq:HHRelsInner} and~\eqref{Eq:HHRelsInf}, that relate the coefficients at the Cauchy horizon and infinity with the $5$ free coefficients $\left\{\bar\phi^{(+)}_0, \bar\phi^{(+)}_1, \bar\psi^{(+)}_0, \bar\psi^{(+)}_1, p_\psi\right\}$. However, it can be proven that the coefficients that enter in the Unruh state conditions~\eqref{Eq:UnruhInnerRegular},~\eqref{Eq:UnruhInfinityRegular}, and~\eqref{Eq:UnruhInfinityVacuum}, namely $\tilde\phi_{0}^{(-)}$, $\tilde\psi_{0}^{(-)}$, $\bar\phi_{-2}^{(\infty)}$, $\bar\psi_{-2}^{(\infty)}$, $\bar\phi_{-1}^{(\infty)}$, $\bar\psi_{-1}^{(\infty)}$, are independent of $\left\{\bar\phi^{(+)}_0,\bar\psi^{(+)}_0\right\}$. Therefore, in order to reproduce the ``true" Unruh state, we would need to satisfy four constraints by adjusting the other three free coefficients. This is sufficient to conclude that the AIEA method, as formulated here with two auxiliary fields and a free parameter $\gamma$, is unable to capture the defining features of the Unruh state.

Previous criticisms to the AIEA method focused on its incapability to reproduce the sub-leading asymptotic contributions of the RSET. On top of these findings, our result provides the most robust counter-argument to date, as we proved that the ASET fails to capture the leading-order behaviour of the RSET in the Unruh state. The implications of this finding cast doubts on the applicability of the AIEA method in general, as its capability to accurately describe semiclassical physics strongly depends heavily on the characteristics of the background spacetime, particularly its symmetries and the number of horizons.

Nonetheless, a minimal extension of this method that incorporates a third auxiliary field described by the action
\begin{equation}
   \mathfrak H(\chi,\chi)=\frac{1}{2}\int\dd[4]x \sqrt{-g} \pqty{-\chi\Delta_4 \chi+F\chi},
\end{equation}
could provide additional integration constants that allows the method to properly describe the ``true'' Unruh state. 
In 4D, there is no fundamental correspondence between states and auxiliary fields, so the decision of how many fields to include should obey a reductionist approach: add as many Weyl-invariant pieces to the anomalous action as necessary to ensure the ASET reproduces (at least qualitatively) the leading-order contributions in the exact RSET, but not more than those. 
Adding more fields would further reduce the discrepancies between the ASET and the RSET, but at the cost of unnecessarily complicating the construction and reducing the method's predictive power. 

\section{Conclusions}
\label{Sec:Conclusions}

The diversity of physical phenomena emerging from the interplay between quantum fields and classical gravity is far from fully explored, particularly in the domain of semiclassical backreaction, which holds potential for revealing new insights into gravitational collapse and stellar equilibrium. In the specific case of gravitational collapse, semiclassical effects in the physical \textit{in} vacuum are anticipated to intensify rapidly near the inner trapping horizon, especially if this approaches the Cauchy horizon over time. This behaviour has been shown to trigger a semiclassical instability, significantly reducing the lifespan of trapped regions~\cite{Barcelo:2020mjw}. The central challenge lies not only in accurately computing the renormalized stress-energy tensor (RSET) but also in estimating its associated backreaction. Advancing this line of research will ultimately demand the development of approximation schemes for the RSET that enable the study of semiclassical phenomena within the framework of modified theories of gravity.

In this work, we examined the anomaly-induced effective action (AIEA) method as a means of approximating the renormalized stress-energy tensor (RSET) within the interiors of charged black holes. Our study specifically assessed the ability of the AIEA method to capture the key features of the Hartle-Hawking and Unruh states near the Cauchy horizon of Reissner-Nordstr\"om spacetimes.

Our findings indicate that, while the AIEA method successfully reproduces the leading-order contributions of the RSET associated with the Hartle-Hawking state, it does not adequately approximate the Unruh state. This limitation arises from the restricted set of boundary conditions that can be imposed within the current AIEA framework. The Unruh state, in particular, requires more stringent constraints on the ASET --- namely, regularity at $\mathcal{H}_{\mathcal{F}}$, constant fluxes at the Cauchy horizon, and emptiness at $\mathscr{I}^{-}$ (equivalent to having Hawking fluxes at $\mathscr{I}^{+}$) --- than can be supported by the available free integration constants in the auxiliary fields. This analysis thus presents a concrete critique of the AIEA method for spacetimes with Cauchy horizons, underscoring the need for alternative approaches or modifications to capture the full scope of semiclassical effects.

We propose that incorporating a third auxiliary field within the AIEA framework may help address some of these limitations. This extension could enable a more accurate approximation of the Unruh state by introducing the additional degrees of freedom necessary to capture both the correct asymptotic behaviours at infinity and near Cauchy horizons.

The current limitations of the AIEA method in accurately modeling the Unruh state in four dimensions highlight a broader challenge within semiclassical descriptions of charged or rotating black holes interiors, where horizon dynamics is unavoidable due to the anticipated classical and semiclassical instabilities at inner horizons. Our findings add to the already existing evidence in favour of inner horizon instabilities and emphasize the crucial role of semiclassical backreaction effects in gravitational collapse. We suggest that future research should then focus on enhancing the AIEA or exploring alternative prescriptions based on effective actions which are capable of approximating both the Hartle-Hawking and Unruh states across a broader spectrum of spacetimes. Steps in generalizing the AIEA method to incorporate the Unruh state in charged black hole spacetimes are currently being explored by the authors and will be reported elsewhere.

\section*{Acknowledgment}

The authors thank Kazumasa Okabayashi for useful conversations. G. Neri thanks Perimeter Institute for its hospitality. This research was supported in part by Perimeter Institute for Theoretical Physics. Research at Perimeter Institute is supported by the Government of Canada through the Department of Innovation, Science, and Economic Development, and by the province of Ontario through the Ministry of Colleges and Universities. J. Arrechea acknowledges funding from the Italian Ministry of Education and Scientific Research (MIUR) under the grant PRIN MIUR 2017-MB8AEZ.

\appendix

\section{$E_{\mu\nu}$, $F_{\mu\nu}$ and $H_{\mu\nu}$ tensors}
\label{App:EFHTensors}

We report here the expressions of the tensors introduced in Eqs.~\eqref{Eq:EFtensor} and~\eqref{Eq:Htensor}. 

\begin{dmath}
    E^{\mu \nu }=4 R^{\mu \rho } R^{\nu }{}_{\rho } \phi - 2 g^{\mu \nu } R_{\rho \sigma } R^{\rho \sigma } \phi - 2 R^{\mu \nu } R \phi + \tfrac {1}{2} g^{\mu \nu } R^2 \phi + 4 R^{\rho \sigma } R^{\mu }{}_{\rho }{}^{\nu }{}_{\sigma } \phi - 2 R^{\mu \rho \sigma \lambda } R^{\nu }{}_{\rho \sigma \lambda } \phi + \tfrac {1}{2} g^{\mu \nu } R_{\rho \sigma \lambda \eta} R^{\rho \sigma \lambda \eta} \phi -  \tfrac {1}{3} \nabla^{\mu }\phi \nabla^{\nu }R -  \tfrac {1}{3} \nabla^{\mu }R \nabla^{\nu }\phi + \tfrac {2}{3} R \nabla^{\mu }\phi \nabla^{\nu }\phi + 2 R \nabla^{\nu }\nabla^{\mu }\phi + \tfrac {14}{3} R^{\mu \nu } \nabla_{\rho }\nabla^{\rho }\phi - 2 g^{\mu \nu } R \nabla_{\rho }\nabla^{\rho }\phi + 2 \nabla^{\nu }\nabla^{\mu }\phi \nabla_{\rho }\nabla^{\rho }\phi -  \nabla^{\nu }\phi \nabla_{\rho }\nabla^{\rho }\nabla^{\mu }\phi -  \nabla^{\mu }\phi \nabla_{\rho }\nabla^{\rho }\nabla^{\nu }\phi -  \tfrac {2}{3} \nabla_{\rho }\nabla^{\rho }\nabla^{\nu }\nabla^{\mu }\phi + \tfrac {1}{3} g^{\mu \nu } \nabla_{\rho }\phi \nabla^{\rho }R + \tfrac {2}{3} \nabla^{\mu }R^{\nu }{}_{\rho } \nabla^{\rho }\phi -  R^{\nu }{}_{\rho } \nabla^{\mu }\phi \nabla^{\rho }\phi + \tfrac {2}{3} \nabla^{\nu }R^{\mu }{}_{\rho } \nabla^{\rho }\phi -  R^{\mu }{}_{\rho } \nabla^{\nu }\phi \nabla^{\rho }\phi -  \tfrac {2}{3} \nabla_{\rho }R^{\mu \nu } \nabla^{\rho }\phi + \tfrac {2}{3} R^{\mu \nu } \nabla_{\rho }\phi \nabla^{\rho }\phi -  \tfrac {1}{3} g^{\mu \nu } R \nabla_{\rho }\phi \nabla^{\rho }\phi + \tfrac {2}{3} \nabla_{\rho }\nabla^{\nu }\nabla^{\mu }\phi \nabla^{\rho }\phi -  \tfrac {10}{3} R^{\nu }{}_{\rho } \nabla^{\rho }\nabla^{\mu }\phi -  \tfrac {4}{3} \nabla_{\rho }\nabla^{\nu }\phi \nabla^{\rho }\nabla^{\mu }\phi -  \tfrac {10}{3} R^{\mu }{}_{\rho } \nabla^{\rho }\nabla^{\nu }\phi -  \tfrac {1}{2} g^{\mu \nu } \nabla_{\rho }\nabla^{\rho }\phi \nabla_{\sigma }\nabla^{\sigma }\phi + \tfrac {1}{3} g^{\mu \nu } \nabla^{\rho }\phi \nabla_{\sigma }\nabla^{\sigma }\nabla_{\rho }\phi + \tfrac {2}{3} g^{\mu \nu } \nabla_{\sigma }\nabla^{\sigma }\nabla_{\rho }\nabla^{\rho }\phi + g^{\mu \nu } R_{\rho \sigma } \nabla^{\rho }\phi \nabla^{\sigma }\phi -  \tfrac {4}{3} R^{\mu }{}_{\rho }{}^{\nu }{}_{\sigma } \nabla^{\rho }\phi \nabla^{\sigma }\phi + 4 g^{\mu \nu } R_{\rho \sigma } \nabla^{\sigma }\nabla^{\rho }\phi - 2 R^{\mu }{}_{\rho }{}^{\nu }{}_{\sigma } \nabla^{\sigma }\nabla^{\rho }\phi -  \tfrac {10}{3} R^{\mu }{}_{\sigma }{}^{\nu }{}_{\rho } \nabla^{\sigma }\nabla^{\rho }\phi + \tfrac {1}{3} g^{\mu \nu } \nabla_{\sigma }\nabla_{\rho }\phi \nabla^{\sigma }\nabla^{\rho }\phi
\end{dmath}

\begin{dmath}
    F^{\mu \nu }=-4 R^{\rho \sigma } C^{\mu }{}_{\rho }{}^{\nu }{}_{\sigma } \phi + R^{\nu \rho \sigma \lambda } C^{\mu }{}_{\rho \sigma \lambda } \phi - 2 R^{\nu \rho \sigma \lambda } C^{\mu }{}_{\sigma \rho \lambda } \phi + R^{\mu \rho \sigma \lambda } C^{\nu }{}_{\rho \sigma \lambda } \phi - 4 C^{\mu \rho \sigma \lambda } C^{\nu }{}_{\rho \sigma \lambda } \phi + 2 R^{\mu \rho \sigma \lambda } C^{\nu }{}_{\sigma \rho \lambda } \phi + \tfrac {1}{2} g^{\mu \nu } C_{\rho \sigma \lambda \eta} C^{\rho \sigma \lambda \eta} \phi + 4 R^{\mu \rho } R^{\nu }{}_{\rho } \psi - 2 g^{\mu \nu } R_{\rho \sigma } R^{\rho \sigma } \psi - 2 R^{\mu \nu } R \psi + \tfrac {1}{2} g^{\mu \nu } R^2 \psi + 4 R^{\rho \sigma } R^{\mu }{}_{\rho }{}^{\nu }{}_{\sigma } \psi - 2 R^{\mu \rho \sigma \lambda } R^{\nu }{}_{\rho \sigma \lambda } \psi + \tfrac {1}{2} g^{\mu \nu } R_{\rho \sigma \lambda \eta} R^{\rho \sigma \lambda \eta} \psi -  \tfrac {1}{3} \nabla^{\mu }\psi \nabla^{\nu }R + \tfrac {2}{3} R \nabla^{\mu }\psi \nabla^{\nu }\phi -  \tfrac {1}{3} \nabla^{\mu }R \nabla^{\nu }\psi + \tfrac {2}{3} R \nabla^{\mu }\phi \nabla^{\nu }\psi + 2 R \nabla^{\nu }\nabla^{\mu }\psi + 2 \nabla^{\nu }\nabla^{\mu }\psi \nabla_{\rho }\nabla^{\rho }\phi + \tfrac {14}{3} R^{\mu \nu } \nabla_{\rho }\nabla^{\rho }\psi - 2 g^{\mu \nu } R \nabla_{\rho }\nabla^{\rho }\psi + 2 \nabla^{\nu }\nabla^{\mu }\phi \nabla_{\rho }\nabla^{\rho }\psi -  \nabla^{\nu }\psi \nabla_{\rho }\nabla^{\rho }\nabla^{\mu }\phi -  \nabla^{\nu }\phi \nabla_{\rho }\nabla^{\rho }\nabla^{\mu }\psi -  \nabla^{\mu }\psi \nabla_{\rho }\nabla^{\rho }\nabla^{\nu }\phi -  \nabla^{\mu }\phi \nabla_{\rho }\nabla^{\rho }\nabla^{\nu }\psi -  \tfrac {2}{3} \nabla_{\rho }\nabla^{\rho }\nabla^{\nu }\nabla^{\mu }\psi + \tfrac {1}{3} g^{\mu \nu } \nabla_{\rho }\psi \nabla^{\rho }R -  R^{\nu }{}_{\rho } \nabla^{\mu }\psi \nabla^{\rho }\phi -  R^{\mu }{}_{\rho } \nabla^{\nu }\psi \nabla^{\rho }\phi + \tfrac {4}{3} R^{\mu \nu } \nabla_{\rho }\psi \nabla^{\rho }\phi -  \tfrac {2}{3} g^{\mu \nu } R \nabla_{\rho }\psi \nabla^{\rho }\phi + \tfrac {2}{3} \nabla_{\rho }\nabla^{\nu }\nabla^{\mu }\psi \nabla^{\rho }\phi + \tfrac {2}{3} \nabla^{\mu }R^{\nu }{}_{\rho } \nabla^{\rho }\psi -  R^{\nu }{}_{\rho } \nabla^{\mu }\phi \nabla^{\rho }\psi + \tfrac {2}{3} \nabla^{\nu }R^{\mu }{}_{\rho } \nabla^{\rho }\psi -  R^{\mu }{}_{\rho } \nabla^{\nu }\phi \nabla^{\rho }\psi -  \tfrac {2}{3} \nabla_{\rho }R^{\mu \nu } \nabla^{\rho }\psi + \tfrac {2}{3} \nabla_{\rho }\nabla^{\nu }\nabla^{\mu }\phi \nabla^{\rho }\psi -  \tfrac {4}{3} \nabla_{\rho }\nabla^{\nu }\psi \nabla^{\rho }\nabla^{\mu }\phi -  \tfrac {10}{3} R^{\nu }{}_{\rho } \nabla^{\rho }\nabla^{\mu }\psi -  \tfrac {4}{3} \nabla_{\rho }\nabla^{\mu }\psi \nabla^{\rho }\nabla^{\nu }\phi -  \tfrac {10}{3} R^{\mu }{}_{\rho } \nabla^{\rho }\nabla^{\nu }\psi - 4 \nabla^{\rho }\phi \nabla_{\sigma }C^{\mu }{}_{\rho }{}^{\nu \sigma } - 4 \nabla^{\rho }\phi \nabla_{\sigma }C^{\mu \sigma \nu }{}_{\rho } - 4 \phi \nabla_{\sigma }\nabla_{\rho }C^{\mu \rho \nu \sigma } -  g^{\mu \nu } \nabla_{\rho }\nabla^{\rho }\phi \nabla_{\sigma }\nabla^{\sigma }\psi + \tfrac {1}{3} g^{\mu \nu } \nabla^{\rho }\psi \nabla_{\sigma }\nabla^{\sigma }\nabla_{\rho }\phi + \tfrac {1}{3} g^{\mu \nu } \nabla^{\rho }\phi \nabla_{\sigma }\nabla^{\sigma }\nabla_{\rho }\psi + \tfrac {2}{3} g^{\mu \nu } \nabla_{\sigma }\nabla^{\sigma }\nabla_{\rho }\nabla^{\rho }\psi + 2 g^{\mu \nu } R_{\rho \sigma } \nabla^{\rho }\phi \nabla^{\sigma }\psi -  \tfrac {4}{3} R^{\mu }{}_{\rho }{}^{\nu }{}_{\sigma } \nabla^{\rho }\phi \nabla^{\sigma }\psi -  \tfrac {4}{3} R^{\mu }{}_{\sigma }{}^{\nu }{}_{\rho } \nabla^{\rho }\phi \nabla^{\sigma }\psi - 2 C^{\mu }{}_{\rho }{}^{\nu }{}_{\sigma } \nabla^{\sigma }\nabla^{\rho }\phi - 2 C^{\mu }{}_{\sigma }{}^{\nu }{}_{\rho } \nabla^{\sigma }\nabla^{\rho }\phi + \tfrac {2}{3} g^{\mu \nu } \nabla_{\sigma }\nabla_{\rho }\psi \nabla^{\sigma }\nabla^{\rho }\phi + 4 g^{\mu \nu } R_{\rho \sigma } \nabla^{\sigma }\nabla^{\rho }\psi - 2 R^{\mu }{}_{\rho }{}^{\nu }{}_{\sigma } \nabla^{\sigma }\nabla^{\rho }\psi -  \tfrac {10}{3} R^{\mu }{}_{\sigma }{}^{\nu }{}_{\rho } \nabla^{\sigma }\nabla^{\rho }\psi 
\end{dmath}

\begin{dmath}
    H^{\mu \nu }=-4 R^{\rho \sigma } C^{\mu }{}_{\rho }{}^{\nu }{}_{\sigma } \psi + R^{\nu \rho \sigma \lambda } C^{\mu }{}_{\rho \sigma \lambda } \psi - 2 R^{\nu \rho \sigma \lambda } C^{\mu }{}_{\sigma \rho \lambda } \psi + R^{\mu \rho \sigma \lambda } C^{\nu }{}_{\rho \sigma \lambda } \psi - 4 C^{\mu \rho \sigma \lambda } C^{\nu }{}_{\rho \sigma \lambda } \psi + 2 R^{\mu \rho \sigma \lambda } C^{\nu }{}_{\sigma \rho \lambda } \psi + \tfrac {1}{2} g^{\mu \nu } C_{\rho \sigma \lambda \eta} C^{\rho \sigma \lambda \eta} \psi + \tfrac {2}{3} R \nabla^{\mu }\psi \nabla^{\nu }\psi + 2 \nabla^{\nu }\nabla^{\mu }\psi \nabla_{\rho }\nabla^{\rho }\psi -  \nabla^{\nu }\psi \nabla_{\rho }\nabla^{\rho }\nabla^{\mu }\psi -  \nabla^{\mu }\psi \nabla_{\rho }\nabla^{\rho }\nabla^{\nu }\psi -  R^{\nu }{}_{\rho } \nabla^{\mu }\psi \nabla^{\rho }\psi -  R^{\mu }{}_{\rho } \nabla^{\nu }\psi \nabla^{\rho }\psi + \tfrac {2}{3} R^{\mu \nu } \nabla_{\rho }\psi \nabla^{\rho }\psi -  \tfrac {1}{3} g^{\mu \nu } R \nabla_{\rho }\psi \nabla^{\rho }\psi + \tfrac {2}{3} \nabla_{\rho }\nabla^{\nu }\nabla^{\mu }\psi \nabla^{\rho }\psi -  \tfrac {4}{3} \nabla_{\rho }\nabla^{\nu }\psi \nabla^{\rho }\nabla^{\mu }\psi - 4 \nabla^{\rho }\psi \nabla_{\sigma }C^{\mu }{}_{\rho }{}^{\nu \sigma } - 4 \nabla^{\rho }\psi \nabla_{\sigma }C^{\mu \sigma \nu }{}_{\rho } - 4 \psi \nabla_{\sigma }\nabla_{\rho }C^{\mu \rho \nu \sigma } -  \tfrac {1}{2} g^{\mu \nu } \nabla_{\rho }\nabla^{\rho }\psi \nabla_{\sigma }\nabla^{\sigma }\psi + \tfrac {1}{3} g^{\mu \nu } \nabla^{\rho }\psi \nabla_{\sigma }\nabla^{\sigma }\nabla_{\rho }\psi + g^{\mu \nu } R_{\rho \sigma } \nabla^{\rho }\psi \nabla^{\sigma }\psi -  \tfrac {4}{3} R^{\mu }{}_{\rho }{}^{\nu }{}_{\sigma } \nabla^{\rho }\psi \nabla^{\sigma }\psi - 2 C^{\mu }{}_{\rho }{}^{\nu }{}_{\sigma } \nabla^{\sigma }\nabla^{\rho }\psi - 2 C^{\mu }{}_{\sigma }{}^{\nu }{}_{\rho } \nabla^{\sigma }\nabla^{\rho }\psi + \tfrac {1}{3} g^{\mu \nu } \nabla_{\sigma }\nabla_{\rho }\psi \nabla^{\sigma }\nabla^{\rho }\psi  
\end{dmath}
\section{Regular states at the event horizon}
\label{App:Regularity Conditions}

We report here the explicit form of the conditions that the auxiliary fields $\phi$ and $\psi$  must satisfy --- in the static, spherically symmetric sector of the solution space --- to produce a regular stress-energy tensor at the event horizon, in terms of the coefficients relative to the following series expansions:
\begin{subequations}
\begin{align}
    \phi&=\sum_{n=0}^\infty \pqty{\frac{r}{r_+}-1}^n \pqty{\bar\phi_n+\tilde\phi_n\ln\abs{\frac{r}{r_+}-1}},\\
    \psi&=\sum_{n=0}^\infty \pqty{{\frac{r}{r_+}-1}}^n \pqty{\bar\psi_n+\tilde\psi_n\ln\abs{\frac{r}{r_+}-1}}.
\end{align}
\end{subequations}
In this appendix, we omit the superscript $(+)$.
As mentioned in the main text (cfr. footnote~\ref{Ftn:HHHorizonRegularity}), regularity imposes $10$ conditions, $5$ of which are redundant for this class of solutions. The remaining conditions form a system of five algebraic equations:
\begin{equation}
\label{Eq:RegSystemHH}
\begin{cases}
    b_{E} \tilde\phi_0^2+2 b_{F} \tilde\psi_0 \tilde\phi_0+\gamma \tilde\psi_0^2=N,\\
    b_{E} \tilde\phi_1^2+2 b_{F} \tilde\psi_1\tilde\phi_1+\gamma \tilde\psi_1^2=0,\\
    A_0\tilde\phi_0 +A_1\tilde\phi_1+B_0\tilde\psi_0 +B_1\tilde\psi_1 =WZ,\\
    C_0\tilde\phi_0+C_1\tilde\phi_1+D_0\tilde\psi_0+D_1\tilde\psi_1=W(b_E\tilde\phi_0\tilde\phi_1+b_F\tilde\phi_0\tilde\psi_1+b_F\tilde\psi_0\tilde\phi_1+\gamma\tilde\psi_0\tilde\psi_1),\\
    E_0\tilde\phi_0 +E_1\tilde\phi_1+F_0\tilde\psi_0 +F_1\tilde\psi_1 =-\epsilon W(b_E\bar\phi_1\tilde\phi_1+b_F\bar\phi_1\tilde\psi_1+b_F\bar\psi_1\tilde\phi_1+\gamma\bar\psi_1\tilde\psi_1)
\end{cases}
\end{equation}
where $\epsilon\equiv 1-\frac{r_-}{r_+}$, and constant coefficients given explicitly by
\begin{alignat*}{2}
    & A_0=6 \left(b_{E} \left(20 \epsilon ^2-12 \epsilon +3\right)+b_{F} \epsilon  \left(8 \epsilon ^2+3\right)\right),\\
    &A_1=\epsilon  \left(b_{E} \left(16 \epsilon ^2-36 \epsilon +9\right)+36 b_{F} (\epsilon -1) \epsilon \right),\\
    &B_0=6 \left(b_{F} \left(20 \epsilon ^2-12 \epsilon +3\right)+\gamma \epsilon  \left(8 \epsilon ^2+3\right)\right),\\
    &B_1=\epsilon  \left(b_{F} \left(16 \epsilon ^2-36 \epsilon +9\right)+36 \gamma (\epsilon -1) \epsilon \right),\\
    &C_0=-18b_F \epsilon(1-\epsilon),\\
    &C_1=3(1-\epsilon)(b_E(3\epsilon-1)+3b_F\epsilon),\\
    &D_0=-18 \gamma \epsilon(1-\epsilon),\\
    &D_1=3(1-\epsilon)(b_F(3\epsilon-1)+3\gamma\epsilon),\\
    &E_0=-36b_F\epsilon(3\epsilon^2-4\epsilon+1),\\
    &E_1=3(b_E(18\epsilon^3-10\epsilon^2+2\epsilon+1)+b_F \epsilon(38\epsilon^2-36\epsilon+9)),\\
    &F_0=-36\gamma\epsilon(3\epsilon^2-4\epsilon+1),\\
    &F_1=3(b_F(18\epsilon^3-10\epsilon^2+2\epsilon+1)+\gamma \epsilon(38\epsilon^2-36\epsilon+9)),
\end{alignat*}
\begin{equation*}
    N=0,\quad W=20 \epsilon ^2-12 \epsilon +3,\quad Z=0.
\end{equation*}

We derive here the possible solutions to the system. The first two equations of the system represent a pair of straight lines in the $(\tilde\phi_0,\tilde\psi_0)$ and in the $(\tilde\phi_1,\tilde\psi_1)$ planes. Solving them gives us four possibilities, depending on the (independent) choices of $\beta_0$ and $\beta_1$ --- defined by the relations $\tilde\phi_0=\beta_0\tilde\psi_0$ and $\tilde\phi_1=\beta_1\tilde\psi_1$ --- among the two angular coefficient values $(-b_F\pm c_\gamma)/b_E$, with $c_\gamma\equiv(b_F^2-b_E\gamma)^{1/2}$. Once a choice is made, the third equation gives
\begin{equation}
    \tilde\psi_1=-\frac{A_0 \beta^i_0+B_0}{A_1 \beta^i_1+B_1}\tilde\psi_0\equiv x^i \tilde\psi_0, \quad i=1,2,3,4,
\end{equation}
and the fourth reduces to
\begin{equation}
    (C_0\beta_0^i+C_1\beta^i_1x^i+D_0+D_1x^i)\tilde\psi_0=W\tilde\psi_0^2 x^i(b_E\beta^i_0\beta^i_1+b_F\beta^i_0+b_F\beta^i_1+\gamma).
\end{equation}
This equation has two different solutions:
\begin{equation}
\label{Eq:fourthequationsolution}
    \tilde\psi_0=0,\quad \tilde\psi_0=\frac{1}{Wx^i}\frac{C_0\beta_0^i+C_1\beta^i_1x^i+D_0+D_1x^i}{b_E\beta^i_0\beta^i_1+b_F\beta^i_0+b_F\beta^i_1+\gamma},
\end{equation}
the second of which does not exist for the choices where $\beta_0=\beta_1$:
\begin{itemize}
    \item If we select the first possibility $\tilde\psi_0=0$, then all four choices collapse to the same solution, that we call \textit{minimal}
    \begin{equation}
    \label{Eq:HHminimal}
    \text{M:}\quad \tilde\phi_0=\tilde\phi_1=\tilde\psi_0=\tilde\psi_1=0.
    \end{equation}
    The last equation trivializes and the system is solved.
    \item If we select the second possibility in~\eqref{Eq:fourthequationsolution} (which exists only for the two choices for which $\beta_0\ne\beta_1$), we can explicitly solve for the tilde coefficients and obtain two solutions
    \begin{equation}
    \text{NM$^\pm$:}\;
    \begin{split}
     &\tilde\phi_0=\frac{3 (1-\epsilon) \left(b_{F}\pm c_\gamma\right) \left(\epsilon  \left(\pm 2 (\epsilon -2) c_\gamma+2 b_{F} (\epsilon +1)+3 \gamma  \epsilon \right)+b_{E} (3 \epsilon -1)\right)}{2 b_{F} \epsilon  \left(8 \epsilon ^2+3\right) \left(b_{F}\pm c_\gamma\right)\pm 2 b_{E} (4 \epsilon  (5 \epsilon -3)+3) c_\gamma-2 b_{E} \gamma  \epsilon  \left(8 \epsilon ^2+3\right)},\\
    &\tilde\phi_1=\frac{9 (1-\epsilon) \left(b_{F}\mp c_\gamma\right) \left(\epsilon  \left(\pm 2 (\epsilon -2) c_\gamma+2 b_{F} (\epsilon +1)+3 \gamma  \epsilon \right)+b_{E} (3 \epsilon -1)\right)}{\epsilon  \left(36 b_{F} (1-\epsilon) \epsilon  \left(b_F\mp c_\gamma\right)\pm b_{E} (4 \epsilon  (4 \epsilon -9)+9) c_\gamma+36 b_{E} \gamma  (\epsilon -1) \epsilon \right)},\\
    &\tilde\psi_0=\frac{3 b_{E} (\epsilon -1) \left(\epsilon  \left(\pm 2 (\epsilon -2) c_\gamma+2 b_{F} (\epsilon +1)+3 \gamma  \epsilon \right)+b_{E} (3 \epsilon -1)\right)}{2 b_{F} \epsilon  \left(8 \epsilon ^2+3\right) \left(b_{F}\pm c_\gamma\right)\pm 2 b_{E} (4 \epsilon  (5 \epsilon -3)+3) c_\gamma-2 b_{E} \gamma  \epsilon  \left(8 \epsilon ^2+3\right)},\\
    &\tilde\psi_1=\frac{9 b_{E} (\epsilon -1) \left(\epsilon  \left(\pm 2 (\epsilon -2) c_\gamma+2 b_{F} (\epsilon +1)+3 \gamma  \epsilon \right)+b_{E} (3 \epsilon -1)\right)}{\epsilon  \left(36 b_{F} (1-\epsilon) \epsilon  \left(b_{F}\mp c_\gamma\right)+b_{E} (4 \epsilon  (4 \epsilon -9)+9) c_\gamma+36 b_{E} \gamma  (\epsilon -1) \epsilon \right)},
    \end{split}
    \end{equation}
    We call these solutions \textit{non-minimal} because they will require a balance between the two auxiliary fields, as dictated by the fifth equation.
    Notice that these solutions are real if and only if $\gamma> \gamma_s$. For $\gamma=\gamma_s$, one necessarily has $\beta_0=\beta_1$, but the fourth and fifth equations trivialize.
\end{itemize}

On the other hand, to study the regularity of solutions that have a linear time dependence, we have to slightly modify the system. Recall that we have to shift $\phi\to\phi+p_\phi t$ and $\psi\to\psi+p_\psi t$ to describe the Unruh state. In this case, we cannot ask for regularity on both branches of the horizon, hence we just demand it for the future branch. This translates into 11 conditions, 6 of which are redundant for this class of solution. The remaining five form a system similar to the one in the previous case~\eqref{Eq:RegSystemHH}:
\begin{equation}
\begin{cases}
    b_{E} \tilde\phi_0^2+2 b_{F} \tilde\psi_0 \tilde\phi_0+\gamma \tilde\psi_0^2=N,\\
    b_{E} \tilde\phi_1^2+2 b_{F} \tilde\psi_1\tilde\phi_1+\gamma \tilde\psi_1^2=0,\\
    A_0\tilde\phi_0 +A_1\tilde\phi_1+B_0\tilde\psi_0 +B_1\tilde\psi_1 =WZ,\\
    C_0\tilde\phi_0+C_1\tilde\phi_1+D_0\tilde\psi_0+D_1\tilde\psi_1=W(b_E\tilde\phi_0\tilde\phi_1+b_F\tilde\phi_0\tilde\psi_1+b_F\tilde\psi_0\tilde\phi_1+\gamma\tilde\psi_0\tilde\psi_1),\\
    E_0\tilde\phi_0 +E_1\tilde\phi_1+F_0\tilde\psi_0 +F_1\tilde\psi_1 =-\epsilon W(b_E\bar\phi_1\tilde\phi_1+b_F\bar\phi_1\tilde\psi_1+b_F\bar\psi_1\tilde\phi_1+\gamma\bar\psi_1\tilde\psi_1)
\end{cases}
\end{equation}
but with a different set of coefficients:\footnote{It may appear that $B_0$ and $B_1$ are modified as well, but, imposing $b_F p_\phi+\gamma p_\psi=0$, we see that this is not the case.}
\begin{gather*}
    A_0=6 \left(b_{E} \left(20 \epsilon ^2-12 \epsilon +3\right) (\hat p_\phi  (\epsilon -1)-1)+b_{F} \left(-3 \hat p_\psi+4 (5 \hat p_\psi-2) \epsilon ^3-32 \hat p_\psi \epsilon ^2+3 (5 \hat p_\psi-1) \epsilon \right)\right),\\
    A_1=\epsilon  \left(b_{E} \left(9 (\hat p_\phi -1)+4 (15 \hat p_\phi -4) \epsilon ^2-36 (\hat p_\phi -1) \epsilon \right)+3 b_{F} \left(3 \hat p_\psi+4 (5 \hat p_\psi-3) \epsilon ^2-12 (\hat p_\psi-1) \epsilon \right)\right),\\
    B_0=6 \left(b_{F} \left(20 \epsilon ^2-12 \epsilon +3\right) (\hat p_\phi  (\epsilon -1)-1)+\gamma  \left(-3 \hat p_\psi+4 (5 \hat p_\psi-2) \epsilon ^3-32 \hat p_\psi \epsilon ^2+3 (5 \hat p_\psi-1) \epsilon \right)\right),\\
    B_1=\epsilon  \left(b_{F} \left(9 (\hat p_\phi -1)+4 (15 \hat p_\phi -4) \epsilon ^2-36 (\hat p_\phi -1) \epsilon \right)+3 \gamma  \left(3 \hat p_\psi+4 (5 \hat p_\psi-3) \epsilon ^2-12 (\hat p_\psi-1) \epsilon \right)\right),\\
    N=b_E\hat p_\phi^2+2b_F\hat p_\phi\hat p_\psi+\gamma\hat p_\psi^2,\\
    Z=-6 (b_{E} \hat p_\phi  (\hat p_\phi  (\epsilon -1)-1)+b_{F} (\hat p_\phi  (2 \hat p_\psi-1) \epsilon -(2 \hat p_\phi +1) \hat p_\psi)+\gamma  \hat p_\psi ((\hat p_\psi-1) \epsilon -\hat p_\psi)),
\end{gather*}
where we rescaled $p_\phi\to\hat p_\phi=\frac{p_\phi}{2\kappa_+}$ and $p_\psi\to\hat p_\psi=\frac{p_\psi}{2\kappa_+}$ to make them dimensionless.

This modification changes the nature of the first equation to a generic conic section, while the second is still a pair of straight lines. Solving the latter is equivalent to choose one of the values for $\beta_1$. In this case, we solve the third equation to get
\begin{equation}
    \tilde\psi_0=\frac{1}{B_0}\pqty{W Z-A_0\tilde\phi_0-(A_1\beta_1+B_1)\tilde\psi_1}.
\end{equation}
Inserting these solutions in the fourth equation reduces it to the form $P_1(\tilde\psi_1)\tilde\phi_0=P_2(\tilde\psi_1)$, where $P_1$ and $P_2$ are respectively a linear and a quadratic polynomial. Barring the case $P_1(\tilde\psi_1)=P_2(\tilde\psi_1)=0$, which can happen only for particular values of $\hat p_\psi$, we solve the equation for $\tilde\phi_0$ and eventually plug everything into the first equation. Despite complicated, one can show that the resulting equation is cubic, and, using the condition for time independence $b_F\hat p_\phi+\gamma\hat p_\psi=0$, that $\tilde\psi_1=0$ is a solution. Corresponding to this, we find the minimal solution
\begin{equation}
    \label{Eq:Uminimal}
    \text{M:}\quad \tilde\phi_0=\hat p_\phi,\quad\tilde\psi_0=\hat p_\psi,\quad\tilde\phi_1=\tilde\psi_1=0,
\end{equation}
which obviously reduces to~\eqref{Eq:HHminimal} when $\hat p_\phi,\hat p_\psi\to 0$. The other two solutions reduce to NM$^\pm$ when the coefficients of time dependence are removed, but they are not worth writing down.

\bibliographystyle{JHEP}
\bibliography{biblio-semiclassical}

\end{document}